\documentclass[usenatbib]{mnras}

\usepackage{graphicx}	
\usepackage{amsmath}	
\usepackage{multicol}
\usepackage{hyperref}
\newcommand{\hi}{H{\sc i} }
\usepackage{ulem, soul}
\usepackage{newtxtext,newtxmath}
\usepackage{anyfontsize}
\usepackage{subcaption}

\title[Intensity Mapping with MIGHTEE-COSMOS]{HI Intensity Mapping with the MIGHTEE Survey: First Results of the \hi Power Spectrum}

\author[A Mazumder et al.]{
Aishrila Mazumder$^{1}$ \thanks{E-mail: aishri0208@gmail.com}, 
Laura Wolz$^{1}$, Zhaoting Chen$^{1,2}$, Sourabh Paul$^{1,3,4}$, Mario G. Santos$^{3,5}$, \newauthor Matt Jarvis$^{6,3}$, Junaid Townsend$^{3}$, Srikrishna Sekhar $^{8,9}$, Russ Taylor $^{9,10,3}$
\\
$^{1}$Jodrell Bank Centre for Astrophysics, Department of Physics and Astronomy, The University of Manchester, Manchester M13 9PL, UK\\
$^{2}$Institute for Astronomy, The University of Edinburgh, Royal Observatory, Edinburgh EH9 3HJ, UK\\
$^{3}$ Department of Physics and Astronomy, University of the Western Cape, Robert Sobukwe Road, Bellville 7535, Cape Town, South Africa\\
$^{4}$ Department of Physics, McGill University, Montreal, QC, Canada H3A 2T8\\
$^{5}$ South African Radio Astronomy Observatory (SARAO), 2 Fir Street, Cape Town, 7925, South Africa\\ 
$^{6}$ Astrophysics, Department of Physics, University of Oxford, Keble Road, Oxford OX1 3RH, UK\\
$^{7}$ Department of Physics and Electronics, Rhodes University, PO Box 94, Makhanda 6140, South Africa\\
$^{8}$ National Radio Astronomy Observatory, 1003 Lopezville Road, Socorro, NM 87801, USA\\
$^{9}$ Inter-university Institute for Data Intensive Astronomy, Department of Astronomy, University of Cape Town, 7701 Rondebosch, Cape Town, South Africa\\
$^{10}$ Department of Astronomy, University of Cape Town, Rondebosch, Cape Town, 7701, South Africa
}

\date{Accepted XXX. Received YYY; in original form ZZZ}

\pubyear{2015}

\begin{document}
\label{firstpage}
\pagerange{\pageref{firstpage}--\pageref{lastpage}}
\maketitle

\begin{abstract}

We present the first results of the \hi intensity mapping power spectrum analysis with the MeerKAT International GigaHertz Tiered Extragalactic Exploration (MIGHTEE) survey. We use data covering $\sim$ 4 square degrees in the COSMOS field using a frequency range 962.5\,MHz to 1008.42\,MHz, equivalent to \hi emission in $0.4<z<0.48$. The data consists of 15 pointings with a total of 94.2 hours on-source. We verify the suitability of the MIGHTEE data for \hi intensity mapping by testing for residual systematics across frequency, baselines and pointings. We also vary the window used for \hi signal measurements and find no significant improvement using stringent Fourier mode cuts. We compute the \hi power spectrum at scales $0.5\,\textrm{Mpc}^{-1} \lesssim k \lesssim 10\,\textrm{Mpc}^{-1}$ in auto-correlation as well as cross-correlation between observational scans using power spectrum domain averaging for pointings. We report consistent upper limits of 29.8 \,mK$^{2}$Mpc${^3}$ from the 2$\sigma$ cross-correlation measurements and 25.82\,mK$^{2}$Mpc${^3}$ from auto-correlation at $k\sim$2\,Mpc$^{-1}$.The low signal-to-noise in this data potentially limits our ability to identify residual systematics, which will be addressed in the future by incorporating more data in the analysis.

\end{abstract}




\section{Introduction}

One of the most important objectives of cosmological observations is measuring the distribution and evolution of dark matter in the Universe. This is traditionally done with large galaxy surveys using their positions as tracers of the underlying matter distribution. With the advent of more sensitive instruments, galaxy surveys are being conducted with increased depth and sensitivity to produce tighter constraints on different cosmological parameters (see, for example \citealt{clustring, des_clustering, des_weak_lens}). Alternatively, the integrated emission of spectral lines from galaxies can also be used to map the large-scale matter distribution. This method is known as line intensity mapping (IM) and is a very efficient technique for tracing the distribution of matter in the Universe (for example \citet{Visbal_2011, kovetz2017lineintensity}). 

Similar to traditional galaxy surveys, IM experiments use biased tracers of dark matter to detect its large-scale distribution. However, unlike galaxy surveys, they do not require the detection of individual galaxies, but rather the cumulative emission of the target spectral line over large voxels in the sky. Thus, it is faster and can achieve good sensitivity with less observing time than galaxy surveys. IM observations are being done with different lines like carbon monoxide CO (e.g. \citealt{CO}), singly ionized carbon C{\sc ii} (e.g.\citealt{CII}), doubly ionized oxygen O{\sc iii} (e.g. \citealt{OIII}), neutral hydrogen \hi (e.g. \citealt{sb_2001}) for providing constraints on the abundance and clustering of gas in the Universe.

The 21-cm hyperfine transition line from neutral hydrogen (\hi) can trace the structure formation out to very high redshifts. Emitted at a rest wavelength of 21.1\,cm (or rest frequency of $\sim$1420\,MHz), it is shifted to lower frequencies due to the expansion of the Universe. This line is detectable using radio telescopes and acts as a valuable cosmological probe at low frequencies. \hi constitutes the bulk of the intergalactic medium (IGM) during the early stages of structure formation, i.e. cosmic dawn (CD). When the first stars and galaxies formed, they emitted ionizing radiation ionizing the neutral IGM - a phase transition period called the epoch of reionization (EoR) \citep{Furlanetto2006}. In the post-EoR era, the pervading ionizing radiation has kept the IGM ionized until the present day, with the \hi gas only confined inside galaxies and haloes. Thus, we can use this single probe to trace back the history of the formation and evolution of structures in the Universe from the present i.e. redshift $z \sim 0$ to the times the first structures were forming i.e. $z\sim$30 \citep{sb_2001, battye_2004, Furlanetto2006, PhysRevLett.100.091303, wyithe, morales2010, marta_theo, zhaoting_theo}. Owing to the unique ability of \hi to trace different structures (i.e. IGM at high redshifts, dark matter halos at lower redshifts) depending on the redshift in question, 21-cm observations have become important science drivers for most radio telescope facilities.

\hi IM can be done using single dish radio telescopes and radio interferometers. At low redshifts, the former probes the larger angular scales or linear scales ($k \lesssim 0.1\,\textrm{Mpc}^{-1}$), while the latter probes smaller angular scales or quasi-linear to non-linear scales ($k \gtrsim 0.1\,\textrm{Mpc}^{-1}$). Forecasts suggest that the SKA Observatory (SKAO), the most sensitive radio telescope to date (currently being built), can make high significance detection of the \hi signal between $z \sim 0.5$ and $z \sim 6$ \citep{red_book}. However, owing to the inherently weak signal, obtaining a detection is extremely challenging, irrespective of the scales of interest. The deterring factors include many orders of magnitude brighter foregrounds, radio frequency interference (RFI), unmodelled systematics and thermal noise. Foregrounds and systematics contaminate cosmological \hi observations irrespective of the redshift targeted. There have been extensive studies on the effect of both foregrounds and systematics for CD/EoR science (for example \citealt{jelic2008,  Datta2010, Chapman2015, barry16, trottwayth2016, ewall2017, aishrila2022}). There have also been studies on how different sources of contamination affect single dish \hi IM experiments (for instance \citealt{wolz_2016, 10.1093/mnras/stx1479, Switzer_2019, steve_2019, steve_2021, sd_data_ch, pb_im}). Effects of different contaminants on \hi IM in the post-EoR era ($z\lesssim\,6$) using interferometers have also been done in a few studies \citep{zchen_1, zchen_ska}. Thus, a large amount of dedicated research is ongoing to better understand the contaminants affecting \hi IM experiments.

 Constraints on the cosmological parameters using single dish telescopes have been produced from the Green Bank Telescope and Parkes. There are detections of power spectrum from \hi cross-correlated with optical galaxy surveys at different redshift regimes: $z<$0.1 \citep{parkes_im}, $z \sim$0.8 \citep{2010Natur.466..463C, Masui_2013, gbt_ii}, $0.6 < z < 1.0$ \citep{wolz_2021}. The MeerKLASS survey uses the SKAO pathfinder MeerKAT interferometer in single dish mode to access the linear cosmological scales \citep{santos2017meerklass}. This poses many challenges for data processing as demonstrated in \citep{wang_meerklass} for the pilot data. Nevertheless, cross-correlating the IM pilot data with WiggleZ Dark Energy Survey \citep{wigglez} has resulted in the detection of the power spectrum \citep{steve_2022}. 

There has also been significant progress in performing \hi IM with interferometers. While there have been upper limits placed on the \hi power spectrum at very high redshift ($7\lesssim z\lesssim 15$) using telescopes like uGMRT, MWA, HERA and LOFAR \citep{pacgia, lofar2, mwa2, hera}, no detections have yet been reported. At lower redshifts (0.78 $\lesssim z \lesssim 2.3$), the Canadian Hydrogen Intensity Mapping Experiment (CHIME, \citealt{chime}) has detected large-scale structures with \hi stacking using Extended Baryon Oscillation Spectroscopic Survey (eBOSS, \citealt{Dawson_2016}) data of galaxies between $0.78\lesssim z \lesssim 1.43$ \citep{chime_galaxy} and Ly$\alpha$ forest \citep{chime_lya} at $z=2.3$. There are also upper limits on the \hi power spectrum using the uGMRT at $1.96 \lesssim z \lesssim 3.58$ \citep{Chakraborty_2021}. However, the most compelling result so far has been achieved with the MeerKAT telescope. Detection of the auto-correlation power spectrum has been reported with a $\sim$1 square degree deep observation (about 96 hours) at two frequency bands centred at redshifts $z\sim0.32$ and $z\sim0.44$ \citep{paul2023detection}.

In this work, we follow up on the MeerKAT detection using data obtained from the MeerKAT International GigaHertz Tiered Extragalactic Exploration (MIGHTEE) survey of the Cosmic Evolution Survey (COSMOS) field. Covering an area of 4 square degrees using multiple pointings and a total observing time of 94.2 hours on source, the data constitutes a small percentage of the full MIGHTEE survey. This data subset is ideal for demonstrating the feasibility of using relatively shallow depth wide area surveys for IM experiments and also provides the first \hi power spectrum upper limits from the MIGHTEE data.

The paper is organised in the following manner: in \autoref{mightee}, we describe the survey and initial data reduction procedure; in \autoref{usability}, we describe the analyses and tests performed to check the usability of the MIGHTEE data for \hi intensity mapping; the results are presented in \autoref{results}; we summarise our findings in \autoref{summary}. Throughout this work, we have used the best fit cosmological parameters from the Planck 2018 results \citep{planck_cosmology}: $\Omega_{\mathrm{M}}$ = 0.31, $\Omega_{\Lambda}$ = 0.68, $\sigma_{8}$ = 0.811, $H_{0}$ = 67.36 km $\mathrm{s}^{-1}$ $\mathrm{Mpc}^{-1}$.

\section{MIGHTEE Data Analysis}
\label{mightee}

The MIGHTEE survey \citep{mightee} is an extragalactic radio survey performed with the MeerKAT telescope \citep{Jonas:2018Jr}. It covers four extragalactic deep fields - COSMOS, XMM-LSS, E-CDFS and ELAIS-S1 using the MeerKAT L-band (900 - 1670 MHz, \citealt{Specs}). The combined area coverage over the four fields is $\sim$20 square degree. MIGHTEE survey has several science goals, including spectral line, continuum and polarization studies. Thus, the fields were chosen to overlap with areas of extensive multi-wavelength data. MIGHTEE fields also have both photometric and spectroscopic data, which are detailed in \citet{rebecca_2020, 10.1093/mnras/stab1956, 10.1093/mnras/staa687, multi_w_cosmos} and also coincide with large surveys like KiDS \citep{Kids}, SDSS \citep{sdss}, GAMA  \citep{gama} and newer ones like DES \citep{des} and DESI \citep{desi}. Given the larger area compared to single-pointing observations and the availability of optical data, the MIGHTEE survey is also suitable for \hi IM studies to measure the power spectrum, both in auto-correlation as well as in cross-correlation with optical studies \citep{sourabh_2021, zchen_1}. \citet{sourabh_2021} show that using the full survey of $\sim$1000 hours, a detection of the \hi signal at z$\lesssim$0.5 is possible with SNR $>$ 7. However, for this work, we use MIGHTEE data of the COSMOS field for auto-correlation studies.

\subsection{Data Processing}
\label{reduction}

\begin{figure}
    \includegraphics[width=\columnwidth, trim={4.4cm 0.20cm 4.4cm 0cm},clip] {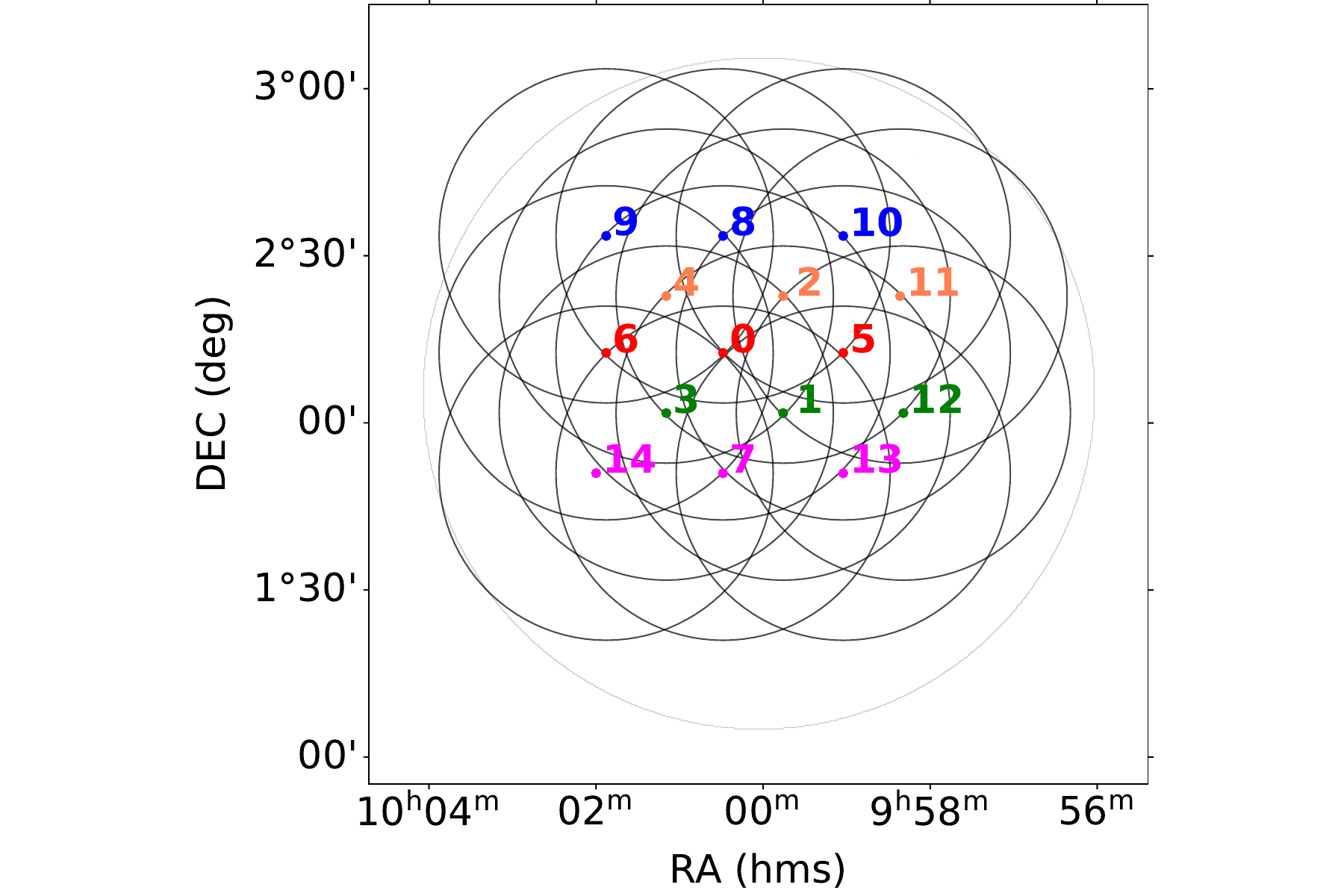}
    \caption{Schematic diagram of the close-packed mosaic observation of MIGHTEE-COSMOS. The black circles represent the FWHM of the MeerKAT L-band primary beam ($\sim$1$^\circ$), with the coloured dots indicating the phase centres of the pointings. The pointing centres are labelled according to their respective COSMOS numbers. Three pointings share a common declination with five different declinations used as phase centres for the MIGHTEE-COSMOS data.}
    \label{cosmos_image}
\end{figure}

Figure \ref{cosmos_image} shows the layout of the MIGHTEE-COSMOS field, with the black circles extending to the primary beam width of the MeerKAT L-band and the phase centres of each of the 15 pointings denoted by the blue dots. The total observing time per pointing is 8 hours, with 6.2 hours on-source and a time resolution of 8 seconds (see \autoref{table:pointings}). Hence, there is 120 hours of total observation time, with 94.2 hours on-source. The pointing centres are arranged in a close-packed mosaic to coincide with the region of the COSMOS field with the best multi-wavelength coverage. For this work, we use the survey data with a frequency resolution of 26\,kHz across 32768 channels in the MeerKAT L-band, primarily used for MIGHTEE-\hi galaxy science \citep{ian_2024}. Early science data from the MIGHTEE survey used 4096 channels \citep{ian_2021}.

The COSMOS pointings cover declinations between +01$^\circ$51\arcmin08.2\arcsec and +02$^\circ$33\arcmin33.8\arcsec. The detailed observation parameters are tabulated in \autoref{table:pointings}. Each group of three pointings have their phase centre at the same declination (specified in the last column of Table \ref{table:pointings}), and for each declination group, the right ascensions are arranged in decreasing order. Throughout the paper, the pointings are arranged identically to Figure \ref{cosmos_image}.
COSMOS\_2 has slightly higher on-source integration than the other pointings; we will use COSMOS\_2 for discussing results for a single pointing.

\begin{table*}
    \caption{Observation details of the MIGHTEE COSMOS Pointings used for this work, arranged in descending order of declination. The constant declination pointings are grouped together with the common declination specified in the last column.}
    \centering
    \begin{tabular}{l|l|l|l|l|l|l}
       ID & Field &  Pointing Center &Date & Time (UTC)& On-source (hr) & Declination (deg)\\
        \hline
        1585928757 & COSMOS\_9 &  (10h01m54s, +02d33m33.79s) & 30.04.2020 & 16:07:25.9-23:36:02.6 & 6.25 & \\
        1585844155 & COSMOS\_8 &  (10h00m29s, +02d33m33.79s) & 02.04.2020 & 16:37:27.3-00:06:03.9 & 6.25 & 2.56$^\circ$\\
        1586016787 & COSMOS\_10 &  (09h59m04s, +02d33m33.79s) & 04.04.2020 & 16:35:26.7-00:04:19.4 & 6.25 & \\
        &&&&& \\
        1621083675 & {COSMOS\_4} &  (10h01m11.063s, +02d22m57.39s) & 15.05.2021 & 13:25:05.08-20:53:58.4 & 6.25 & \\
        1619963656 & {COSMOS\_2} &  (09h59m46.15s, +02d22m57.39s) & 02.05.2021 & 14:17:24.7-21:46:17.3 & 6.40 & 2.38$^\circ$\\
        1586188138 & {COSMOS\_11} &  (09h58m21s,+02d22m57.39s) & 06.04.2020 & 16:10:28.0-23:39:04.6 & 6.25 & \\
        &&&&& \\
        1585498873 & {COSMOS\_6} &  (10h01m54s, +02d12m20.99s) & 29.03.2020 & 16:43:19.7-00:11:48.4 & 6.25 & \\
        1587911796 & {COSMOS\_0}$^{\dagger}$ & (10h00m28.6s, +02d12m21s) & 26.04.2020 & 14:58.02.3-22:26.38.9 & 6.25 & 2.21$^\circ$\\
        1585413022 & {COSMOS\_5} &  (09h59m04s, +02d12m20.99s) & 28.03.2020 & 16:52:28.9-00:21:13.5 & 6.25 & \\
        &&&&& \\
        1622376680 & {COSMOS\_3} &  (10h01m11.053s, +02d01m44.60s) & 30.05.2021 & 12:34:21.9-20:03:22.5 & 6.25 & \\
        1617809470 & {COSMOS\_1} & (09h59m46.15s, +02d01m44.60s) & 07.04.2021 & 15:53:37.9-23:22:22.5 & 6.25 & 2.03$^\circ$\\
        1586274966 & {COSMOS\_12} &  (09h58m21s, +02d01m44.60s) & 07.04.2020 & 16:17:33.1-23:46:17.7 & 6.25 & \\
        &&&&& \\
        1586791316 & {COSMOS\_14} &  (10h01m53s, +01d51m08.20s) & 13.04.2020 & 15:44:06.4-23:12:43.0 & 6.25 & \\
       1585671638 &  {COSMOS\_7} &  (10h00m29s, +01d51m08.20s) & 31.03.2020 & 16:42:45.4-00:11:30.0 & 6.25 & 1.85$^\circ$\\
        1586705155 & {COSMOS\_13} &  (09h59m04s, +01d51m08.20s) & 12.04.2020 & 15:48:05.2-23:16:57.8 & 6.25 & \\
        \hline
    \end{tabular}
\flushleft{$^{\dagger}$This pointing is originally labelled COSMOS, but has been relabelled here as COSMOS\_0 for consistency.}
    \label{table:pointings}
\end{table*}

The observations used here consist of interleaved target scans of around an hour between calibrator observations for each pointing, giving seven scans per pointing. The combined observing time of the MIGHTEE-COSMOS field is similar to the observing time for the first detection of \hi power spectrum using MeerKAT observation of the DEEP2 field \citep{paul2023detection}. However, DEEP2 consisted of a single deep pointing instead of multiple pointings used here. The total area covered by this close-packed mosaicing strategy is $\sim$4 square degrees.  

The data processing steps are described in detail in \citet{ian_2024} and are briefly outlined here for completeness. The data has a high native frequency resolution of 26\,kHz, and is primarily observed for MIGHTEE-\hi galaxy science. The target visibilities are Level-1 (L1) calibrated visibilities recovered from KAT Data Access Library (\texttt{KATDAL}\footnote{\url{https://github.com/ska-sa/katdal}}). These L1 visibilities have reference calibration solutions applied to the targets using the SARAO Science Data Processor (SDP, \citealt{sdp}). The bandpass solutions are derived from the primary calibrator (J0408-6545). The time-dependent complex gain solutions are from the secondary calibrator (J1008+0740). The same calibrator sources are used across all pointings.

The calibration report for the L1 visibilities for each pointing can be obtained from the SARAO archive\footnote{\url{https://archive.sarao.ac.za}} using the observation IDs listed in \autoref{table:pointings}. We summarise the important details here. The initial flags remove known RFI-contaminated channels (see \citealt{flags} for details).  Then, the calibration pipeline performs further flagging to remove time steps and baselines affected by RFI. For the channels of our interest (i.e. between 962.5\,MHz to 1008.4\,MHz), the RFI flagging step usually results in a flagging percentage $\lesssim$25\%. The subsequent calibration steps use antenna m060 as the reference antenna for deriving solutions. The delay and bandpass calibration is performed in the frequency range 973-1000\,MHz using the primary calibrator (J0408-6545).  
Inspection of the solutions shows them to be  stable and smooth for the COSMOS pointings. J1008+0740 i.e. the secondary calibrator is used to derive the time-dependent complex gain solutions. These solutions were applied to the targets i.e. the COSMOS fields and retrieved these calibrated data as measurement sets. This data is subsequently split into two sub-bands-  \texttt{LOW}(960-1150\,MHz, channels averaged to 104.5\,kHz resolution) and \texttt{MID} (1290–1520\,MHz, channels kept at native resolution of 26.1\,kHz).

These calibrated visibilities are further processed in the following steps :
\begin{itemize}
    \item First round of flagging using \texttt{TRICOLOUR} package \citep{tricolor}. It uses the \texttt{SUMTHRESHOLD} algorithm \citep{offringa2010} to clip values where the sum within a time step and channel exceeds a threshold calculated iteratively. This round flags residual low-level features.    
    \item Construction of a deconvolution mask from the deep continuum image as a part of MIGHTEE continuum science \citep{ian_2024, hale_2024}.
    \item Image plane deconvolution using \texttt{WSCLEAN} to create an image of size 10240$\times$10240 and pixel size 1.1\arcsec with robust weighting (r=-0.3).
    \item \texttt{CLEAN} component model obtained from the previous step is interpolated in frequency onto a higher number of sub-bands (909 for \texttt{LOW} band used here) to create a spectrally smooth model.
    \item The \texttt{CLEAN} components of the frequency-dependent model images are used in the \texttt{WSCLEAN} predict mode to create model visibilities for each sub-band.
    \item One round of phase and delay self-calibration with 32s time interval and entire bandwidth for each of the \texttt{MID} and \texttt{LOW} sub-bands as frequency interval. 
    \item Visibility domain continuum subtraction by applying the best-fit gain solutions from the self-calibration step to the data and subtracting the model visibilities. 
    \item Another round of flagging on residual data.
\end{itemize}

Each pointing of the COSMOS data undergoes this process to produce the final calibrated visibilities used here. The MIGHTEE \hi galaxy detection analysis involves additional imaging steps described in \citet{ian_2024}. However, our approach to intensity mapping analysis does not require imaging. Throughout the rest of the paper, we use these continuum-subtracted visibilities. We are interested in the redshift range of $0.2 \lesssim z \lesssim 0.5$, thus we use the \texttt{LOW} band for this work. 

\subsection{Visibility Domain Gridding}
\label{uv-grids}
\begin{figure}
\centering
\includegraphics[width=\columnwidth, trim={0.8cm 0cm 2.5cm 0cm}, clip]{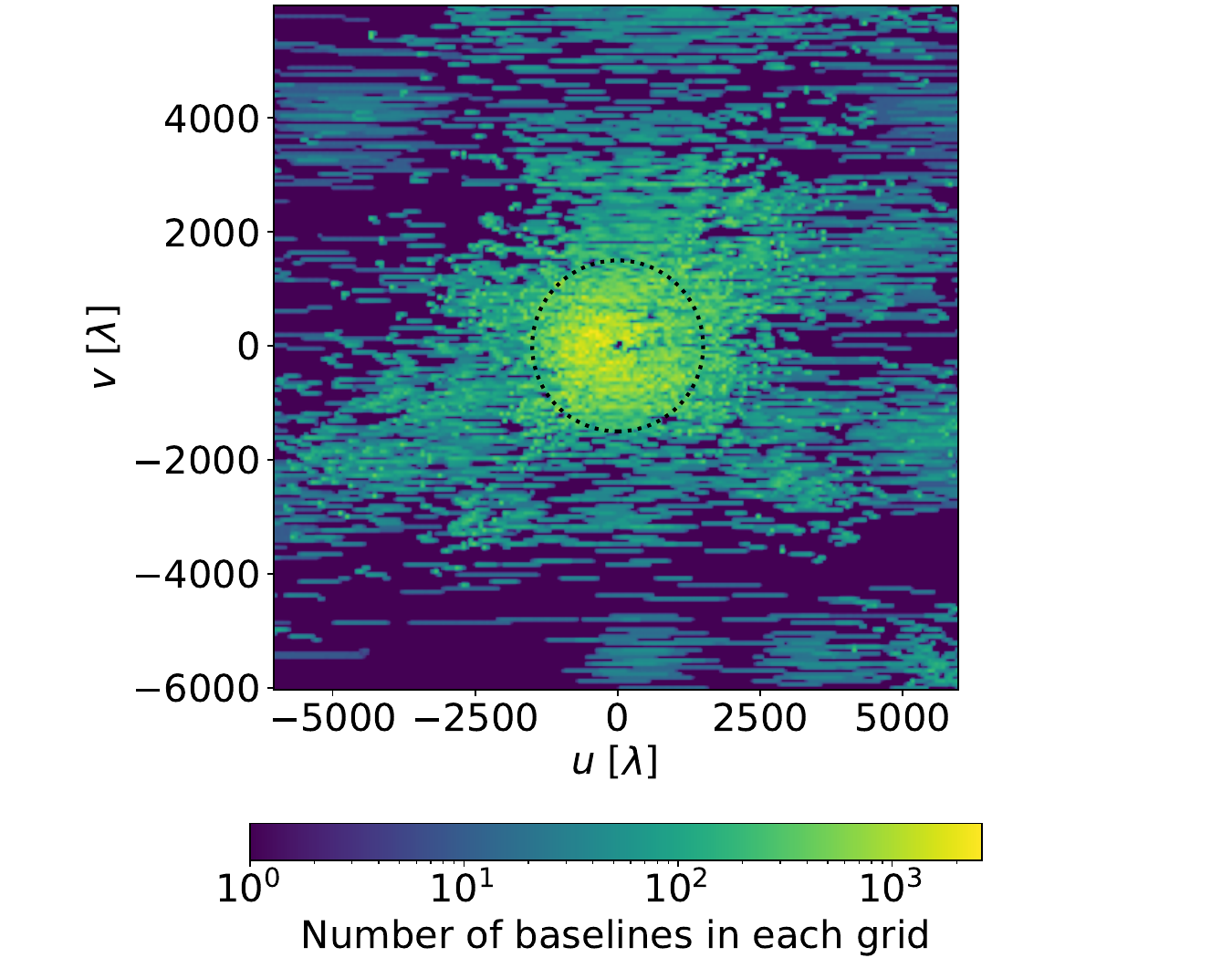}
\caption{Gridded uv plane for the full on-source observation track of COSMOS\_2. Each cell has a size u=v=60$\lambda$. The black dashed circle is the zone between $\pm$1000$\lambda$, used for the PS analysis.}
\label{counts}
\end{figure}

The MIGHTEE-COSMOS data consists of continuous target scans of about an hour each, totalling $\sim$6 hours on-source tracks per pointing. For each pointing, we grid the $uv$-plane into discrete cells to generate gridded visibility cubes in the ($u,v, \nu$), i.e. baseline-frequency domain. Each grid has size $\Delta u= \Delta v=60\lambda$ (where $\lambda$ corresponds to the central wavelength). The gridding is performed as a function of frequency, with each visibility assigned a single ($u,v, \nu$) cell. This gridding approach can change the ($u,v$) cell assigned to each visibility as a function of frequency. The visibilities include only the baselines with $\lesssim$20\% flagged data. The chosen cell size (calculated at the band centre) makes the primary beam width in $\textbf{\textit{k}}_{\perp}$ space negligible, thereby minimising the effect of mixing different $\textbf{\textit{k}}_{\perp}$ modes. Each scan, i.e. interleaved time steps between calibrator observations, is gridded separately for each pointing.
Figure \ref{counts} shows the distribution of the visibilities in the gridded $uv$ plane for a particular pointing (COSMOS\_2) for the full on-source observation. It is evident from Figure \ref{counts} that the \textit{uv}-cells are not sampled with a uniform density far away from the centre of the plane. MeerKAT has more short baselines than long ones. Thus, across an observing session, the sampling is sparser as one goes away from the centre of the $uv$ plane. The observed fields being located at declinations that remain above the horizon for less time than those further south\footnote{For details on track length as a function of declination, see \citealt{Specs}} also contribute to the sparse sampling. Additionally, the individual sources are resolved at long baselines, making the Poisson sampling noise dominant over the clustering signal. All these effects combined justify excluding long baselines for the \hi power spectrum measurements.  

Grid points shown by the black dashed circle in Figure \ref{counts} fall within $\pm$1000$\lambda$ from the centre of the \textit{uv}-plane (corresponding to $k_\perp$=10 Mpc$^{-1}$). This region is uniformly dense for all the pointings. Thus, using only visibilities within a uv-distance of 1000$\lambda$ results in a uniform sampling of $k_\perp$ modes in the final power spectrum estimates. The various analyses performed in the power spectrum domain throughout the paper utilise the gridded visibilities described here.

\subsection{Power Spectrum Estimator}
\label{estimator}
In this work, we use the calibrated MIGHTEE-COSMOS data to estimate the \hi power spectrum. Each pointing has a different phase centre with a relatively low signal-to-noise, and we average over all the pointings, which reduces the noise to enable a combined \hi power spectrum measurement. In this analysis, we incoherently average the power spectra measurement in Fourier space. This is different from coherent averaging in the visibility domain done for deep integration over a single pointing such as the DEEP2 analysis \citep{paul2023detection}. In this section, we describe its formalism.  For the remainder of the article, unless stated explicitly, averaged or combined power spectrum refers to the power spectrum estimated from incoherently averaged data. 

To calculate the power spectrum, the calibrated interferometric visibilities $V(\textbf{\textit{b}},t,\nu)$ are gridded as described in Section \ref{uv-grids}. We then Fourier transform the gridded cube along the frequency axis, resulting in gridded visibility in the Fourier domain, i.e. $\tilde{V}(\textbf{\textit{b}},t,\eta)$. This process is similar to the `delay spectrum' approach \citep{Morales_hewitt, Parsons2012} of calculating the power spectrum from the visibilities. The gridded visibility cube linearly combines data from multiple baselines in the power spectrum estimation process (a similar method is also used for imaging before Fourier transforming visibilities). Thus, following the convention in \citet{morales_2019}, we estimate the ``reconstructed" power spectrum in this work.

The Fourier conjugate of the frequency,  $\eta$ is related to $k_{\parallel}$, i.e. Fourier modes along the line-of-sight via $k_{\parallel} = \frac{2\pi \eta H(z)}{\lambda_{21}(1+z)^2}$, where $\lambda_{21}$ is the rest-frame wavelength of the 21-cm spin-flip transition line, H($z$) is the Hubble parameter corresponding to the redshift $z$ at the central observing frequency $\nu$\footnote{The redshift $z$ is given by $1+z = \nu_{21}/\nu$}. The angular Fourier modes (perpendicular to the line of sight) $\textbf{\textit{k}}_{\perp}$ is related to the baseline vector \textbf{\textit{b}} as $\textbf{\textit{k}}_{\perp} = \frac{2\pi\textbf{\textit{b}}}{\lambda X}$, where $\lambda$ is the wavelength corresponding to $\nu$ and $X$ is the co-moving distance to the corresponding redshift. 

The observed visibilities for a patch of sky with the same $\textbf{k}_{\perp}$ at a certain frequency should have the same value (plus a thermal noise component) at any given time. Hence, in the Fourier domain, $\tilde{V}(\textbf{\textit{b}},t,\eta)$ can be written as $\tilde{V}(\textbf{\textit{k}})$, with $\textbf{\textit{k}} = (\textbf{\textit{k}}_{\perp},k_{\parallel})$. The corresponding 3D power spectrum $P_D$ is given by \citep{Morales_hewitt, Parsons2012}:

\begin{equation}
    P_D(\textbf{\textit{k}}_{\perp}, k_{\parallel}) = \Big(\frac{X^{2}Y}{\Omega_{ps} B}\Big)\Big(\frac{\lambda^{2}}{2k_{B}}\Big)^{2} \operatorname{Re}\{\tilde{V_i}(\textbf{\textit{k}}_{\perp},k_\parallel) \tilde{V_j}^*(\textbf{\textit{k}}_{\perp},k_\parallel)\}  
    \label{2dps_eq}
\end{equation}

where $k_{B}$ is the Boltzmann constant, $Y$ represents the comoving depth along the line-of-sight corresponding to bandwidth $B$. $\Omega_{ps}$ is the power squared primary beam, defined as $\Omega_\mathrm{ps}=\int dl dm |A(l,m)|^2$, with $A(l,m)$ being the primary beam pattern, as opposed to the standard integrated power of the beam obtained by an integration over the beam i.e. $\int dl dm A(l,m)$ (see \citealt{Parsons_2014} for details).

For each pointing, we can estimate the power spectrum either as auto-correlation or as cross-correlation between different observational scans. For the auto-correlation, we compute the gridded visibilities via $\tilde{V_j} = \tilde{V_i^*}$ in Equation \ref{2dps_eq}. The auto-correlation power spectrum contains an additive thermal noise bias term along with the observed sky signal. However, we can infer the noise in the data using observation parameters and remove it from the auto-power spectrum to estimate the noise-free signal power spectrum. In the case of cross-correlation, we create two visibility sets from the observational scans. For example, we can create $\tilde{V_i}$ and $\tilde{V_j}$ from alternate 1-hour scans. They are split into ``even scans" and ``odd scans" based on the parity of the respective scan numbers, i.e. scan numbered 0 is even, scan numbered 1 is odd and so on. The cross-correlation removes the uncorrelated noise bias from the independent samples, in principle measuring the signal power spectrum. 
We conducted tests on the impact of the timescales for creating the cross-correlation samples, and we could not identify significant differences in the cross-correlation amplitudes within the limited signal-to-noise of our observations.  
The 3D powers from both auto and cross-correlations are binned into cylindrical or spherical bins to give the corresponding power spectra. Each pointing of MIGHTEE-COSMOS observes a slightly different part of the sky, so the 3D powers obtained in Equation \ref{2dps_eq} are averaged over all the pointings to get the final averaged power spectrum.

For incoherent averaging, we follow the steps below:
\begin{itemize}
    \item Fourier transform the gridded visibilities along the frequency axis per pointing to obtain the 3D  power spectrum
    \item Weight the 3D $k$ cell by the respective baseline density for each pointing. The weighting accounts for the differences in baseline coverage among the pointings
    \item Average all weighted 3D power spectra to obtain the averaged 3D power
    \item Bin the averaged 3D power spectrum into spherical/cylindrical bins 
\end{itemize}

The final combined 3D power spectrum is:

\begin{equation}
    P_{D} (\textbf{\textit{k}}_{\perp}, k_{\parallel}) =\frac{\sum\limits_{i=1}^{15} w_{i}(\textbf{\textit{k}}_{\perp}, k_{\parallel})\,P_D^i(\textbf{\textit{k}}_{\perp}, k_{\parallel})}{\sum\limits_{i=1}^{15} w_{i}(\textbf{\textit{k}}_{\perp}, k_{\parallel})} 
    \label{ic_eq}
\end{equation}

where the $i$ is over all 15 pointings and $P_D^i(\textbf{\textit{k}}_{\perp}, k_{\parallel})$ is the 3D power for the $i$th pointing, weighted by the factor $w_{i}$ which is the baseline density per 3D $k$ pixel. 

The 2D cylindrical power spectrum, $P(k_{\perp}, k_{\parallel})$ is calculated for the MIGHTEE COSMOS data binning the averaged 3D power spectrum, $P_{D}(\textbf{\textit{k}}_{\perp}, k_{\parallel})$ into cylindrical bins in the line-of-sight and transverse Fourier modes, weighted by the inverse noise variance giving the estimator:

\begin{equation}
    P{^i}(\textbf{\textit{k}}_{\perp}, k_{\parallel}) = \frac{\Big(\Sigma_{j} P_{D}^j(\textbf{\textit{k}}_{\perp}, k_{\parallel}) \, s^{iv}_{j}(\textbf{\textit{k}}_{\perp}, k_{\parallel})\Big)}{\Big(\Sigma_{j}s^{iv}_{j}(\textbf{\textit{k}}_{\perp}, k_{\parallel})\Big)}
    \label{inv_var_eqn}
\end{equation}

where the weight $s^{iv}_{j}$ is the inverse noise variance $1/\sigma^2_j(\textbf{\textit{k}}_{\perp},k_{\parallel})$, $j$ loops over all the $k$ modes in the $i$th $k$ bin.  Owing to the differences in observing days and pointing centres, we simulate the noise variance for each pointing separately (discussed in Section \ref{noise_sim}). The 3D power spectrum is spherically averaged to obtain the statistically relevant 1D or spherical power spectrum using Fourier modes outside the foreground-dominated region or the so-called foreground wedge \citep{Datta2010}.

\subsubsection{Foreground Avoidance}
For a sky with pure statistically isotropic and homogeneous \hi signal, the power spectrum should be a function of $k=\sqrt{{k}_{\perp}^2+k_{\parallel}^2}$. However, real observations will have contributions from systematics and thermal noise (which is highly dependent on the baseline distribution for the particular observation). The isotropy of \hi signal is also broken by redshift space distortions. This effect reduces the amplitude of fluctuations at large ${k}_{\parallel}$ at the small spatial scales observed by an interferometer. Astrophysical foregrounds are spectrally smooth, their interaction with the instrument chromaticity concentrates power at small ${k}_{\parallel}$ or short delays creating the foreground wedge. In interferometric observations, the wedge allows us to isolate the foregrounds from the \hi signal using foreground avoidance \citep{Datta2010, Morales2012, Trott2012, Parsons2012, Vedantham2012, hera, paul2023detection}. Foreground avoidance, in essence, excludes the foreground dominated ($\textbf{\textit{k}}_{\perp}$, ${k}_{\parallel}$) modes inside the wedge for calculating the spherical power spectrum. This method utilises data in the  foreground-free region and does not require additional foreground removal. This work uses foreground avoidance to measure the final spherically averaged power spectrum $P(k)$. 

For an instrument with a primary beam size $\theta_{B}$, foregrounds should ideally be confined within the "horizon limit" \citep{liu2014}, given by:

\begin{equation}
    k_{\parallel} = \frac{XH(z)\mathrm{sin\theta_{B}}}{c(1+z)}{k}_{\perp}
    \label{hori}
\end{equation}
where $H(z)$ is the Hubble parameter. Using Equation \ref{hori} for calculating the primary beam limit, we get $k_{\parallel}\sim 0.01k_{\perp}$ for the MeerKAT frequency band used here. However, due to imperfections in the instrument, the presence of calibration errors and leakage from beyond the instrument beam, the contamination can extend beyond these modes and start affecting higher modes as well \citep{Datta2010, aishrila2022}. Thus, we employ the conservative horizon limit, i.e. $\mathrm{sin\theta=1}$, giving ${k}_{\parallel} \sim 0.3 {k}_{\perp}$. We also use different $k$ mode cuts (discussed in detail in Section \ref{cyl_p}) to test for improvements in our final measurements by excluding more potentially contaminated modes. 

The uncertainties in the calculated power for each $k$ mode of the final power spectrum are calculated from the variance of the power in each bin as:

\begin{equation}
    (\Delta P_D^i(\textbf{\textit{k}}_{\perp}, k_{\parallel}))
    =\sqrt{\frac{\Big(\sum\limits_{j} (P_D^j(\textbf{\textit{k}}_{\perp}, k_{\parallel}) - P_D^i(\textbf{\textit{k}}_{\perp}, k_{\parallel}))^2 s_j^2(\textbf{\textit{k}}_{\perp}, k_{\parallel})\Big)}{\Big(\sum\limits_{j}s_j(\textbf{\textit{k}}_{\perp}, k_{\parallel})\Big)^2}} 
    \label{err_eq}
\end{equation}
where the index $j$ loops over all modes falling in the $i$th k-bin.  

\subsubsection{Noise in Incoherent Averaging}
As described above, the power spectrum measured with incoherent averaging is the so-called "square-then-average" method as opposed to the coherent averaging or the "average-then-square" method \citep{Liu_2016}. The noise scales down slower in the former method compared to the latter. Roughly, for N pointings, the noise amplitude scales down by 1/$\sqrt{N}$, since N independent measurements are averaged after squaring (as opposed to averaging N independent realizations and then squaring, which reduces the noise by 1/N). Thus, for $\sim$100 hours of data over 15 pointings, the noise is equivalent to $\sim$25 hours or $\sim$4 times that of a single 100-hour pointing. Thus, despite an on-source time of 94.2 hours for MIGHTEE-COSMOS (similar to $\sim$96 hours for DEEP2 used in \citet{paul2023detection}), the expected signal-to-noise is that of $\sim$24 hours of integration. Using the radiometer equation, for the used bandwidth of 46\,MHz, we expect a root-mean-squared thermal noise fluctuation of $\sim$5\,mK for a single pointing of 94.2 hours and about 10\,mK if we divide the time into 15 pointings. Simulations from \citet{sourabh_2021} show that using the full $\sim$1000 hours of the MIGHTEE survey, \hi intensity mapping power spectrum detection can be obtained even by incoherent averaging. In this work, we use a part of the full survey data and obtain the first upper limits from MIGHTEE observations.

\section{Quality Assessment for MIGHTEE Data}
\label{usability}

The redshifted 21-cm signal is an inherently weak signal and prone to contamination by several factors. The data calibration, as described above, invariably leaves some residual systematics and possibly, low-level RFI features. The MIGHTEE-COSMOS pointings are spread over a large area ($\sim$4 square degree), with possibilities for declination-dependent systematic contamination. In this section, we present the data quality assessments to check for such effects for individual pointings and the combined data.

\subsection{Residual RFI}
\label{vis_data}

\begin{figure}
\centering
\includegraphics[width=\columnwidth]{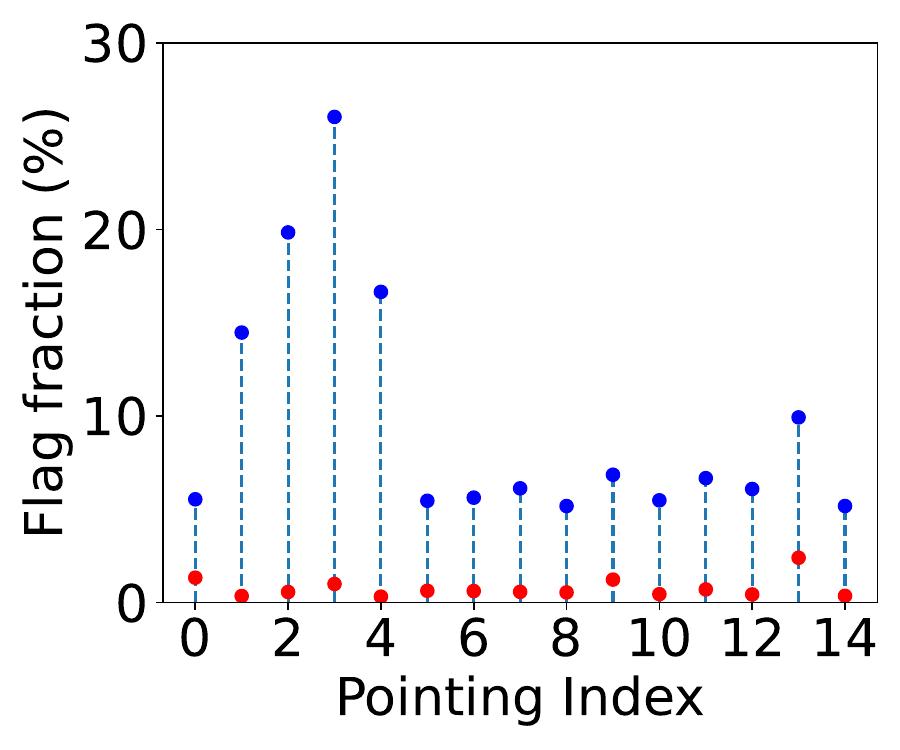}
\caption{Flagging fraction of individual pointings over the entire \texttt{LOW} band (blue points). The overall flagging percentage is  $<$30\% for all 15 pointings. The red points are plotted for a smaller sub-band corresponding to a frequency range 962.5-1008.4\,MHz, with an additional criterion of excluding baselines with more than 20\% flagging. This results in $<$10\% flagging for all cases.}
\label{flags_stem}
\end{figure}

MeerKAT L-band data has several frequency channels pre-flagged by SARAO since they are affected heavily by known RFI contamination \citep{flags}. The extra flagging steps described in \autoref{reduction} remove additional RFI contaminants. As a first check for data quality, the total flagging percentage for the entire \texttt{LOW} band as a per pointing is shown by blue points in Figure \ref{flags_stem}. We see that the percentage of flagged data is below 30\% for all pointings using the full \texttt{LOW} band, with the highest flagging percentage in COSMOS\_3 ($\sim$27\%). The overall trend across most of the pointings shows $\lesssim$15\% flagging, with the exceptions COSMOS\_1, COSMOS\_2, COSMOS\_3, COSMOS\_4 and COSMOS\_13. Since there are multiple pointings at the same declination, we checked for any declination-dependent variation in the flagging percentage. There is no indication of such effects from the flagging percentages since COSMOS\_2, COSMOS\_4 and COSMOS\_11 are at 2.38$^\circ$, but the latter has $\sim$8\% data flagged. Similarly, COSMOS\_1 and COSMOS\_3 show higher flagging percentages than COSMOS\_12 ($\sim$6\%); but all three are at a declination of 2.03$^\circ$. The same is also true for COSMOS\_13 when compared to COSMOS\_7 and COSMOS\_14 at 1.85$^\circ$. 

\begin{figure*}
\centering
\includegraphics[width=\textwidth, height=15cm]{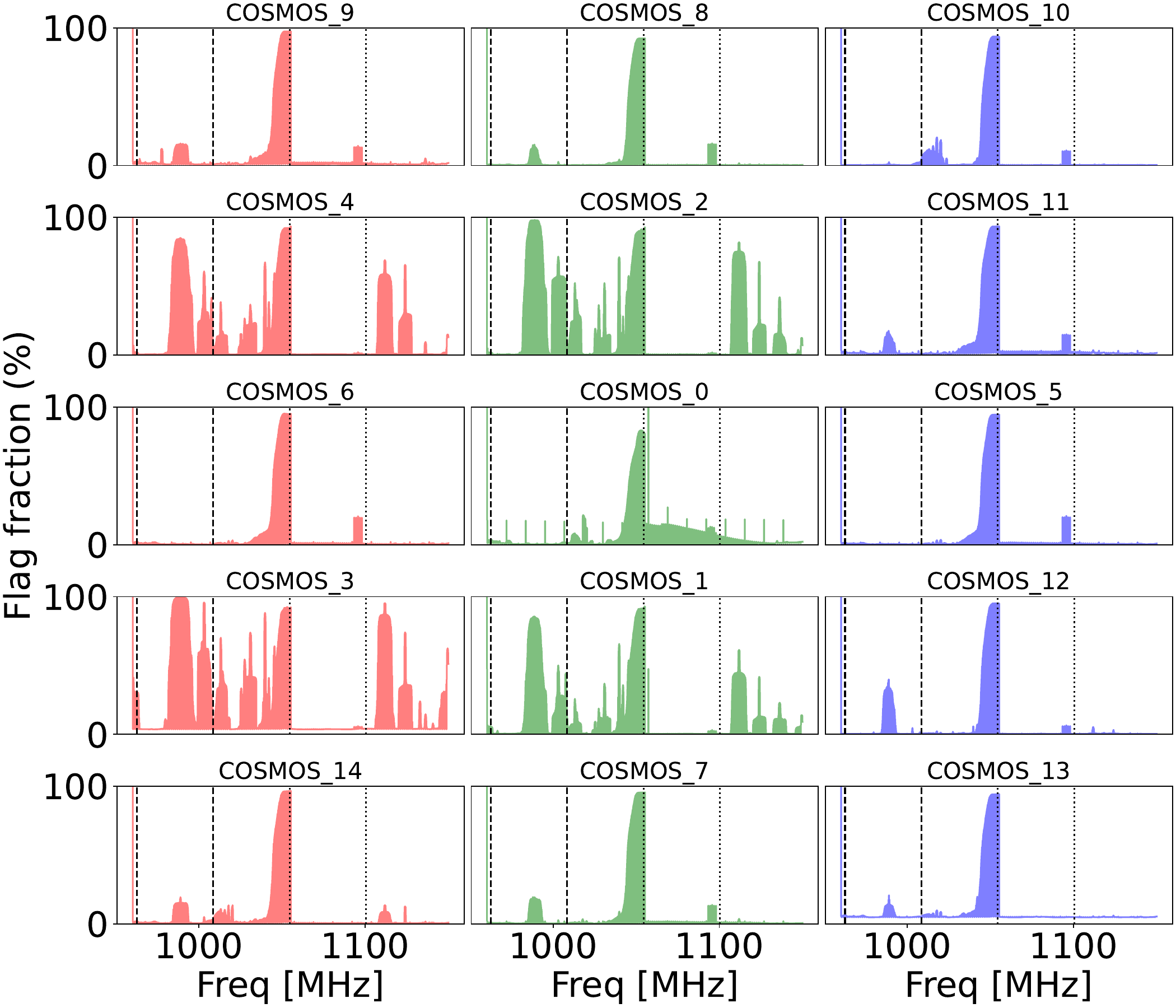}
\caption{Flagging fraction of individual pointings as a function of frequency of the full \texttt{LOW} band. The region enclosed by black dashed lines is the frequency range for $z\sim0.44$, and the black dotted line is that for $z\sim0.32$ used in \citet{paul2023detection}. Panels in each row represent pointings at the same declination. The columns are arranged from left to right in order of right ascensions as shown in Figure \ref{cosmos_image} and rows are in descending order of declination.}
\label{flags_chan}
\end{figure*}

For further investigation, we plot the flagging percentage as a function of frequency for each pointing. Figure \ref{flags_chan} shows the percentage of data flagged per channel of the \texttt{LOW} band for the pointings as indicated in each panel. Constant declination pointings are grouped in rows. In \citet{paul2023detection}, the cleanest part of the band was obtained between 962.55-1008.42\,MHz (centred at $\sim$986\,MHz) for redshift $\sim$0.44 and between 1054-1100\,MHz, i.e. centred at $\sim$1077.5\,MHz for z$\sim$0.32. The respective regions are marked by dashed and dotted lines in Figure \ref{flags_chan}. The higher frequency sub-band is reasonably clean for all the MIGHTEE pointings, but the lower sub-band shows large flagging fractions in some cases, e.g. for pointings COSMOS\_1, COSMOS\_2, COSMOS\_3, and COSMOS\_4. Ambient factors like the presence of the sun or observations taken during local sunset (or sunrise) can result in additional features leading to excess flagging. However, many of the pointings have the sun present, while  COSMOS\_13 and COSMOS\_14  also cover local sunset but do not show such high amounts of flagging. Further investigations show that the four pointings were taken a year after the others, hinting at the presence of some excess ambient interference leading to more flagging for these specific observations.

\begin{figure*}
\centering
\includegraphics[width=\textwidth, height=15cm]{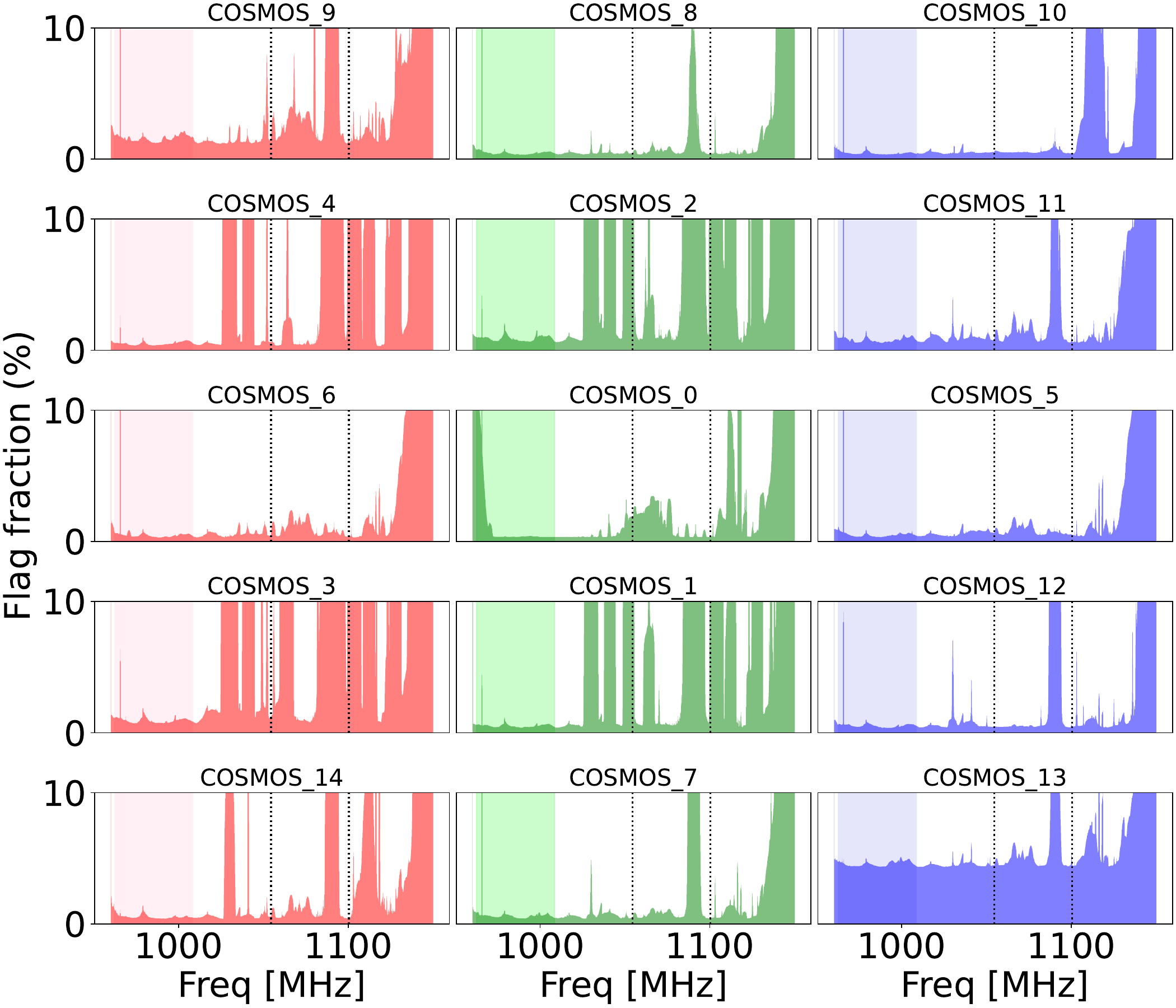}
\caption{Flagging fraction of individual pointings as a function of the frequency of the \texttt{LOW} band excluding all baselines which are $\gtrsim$20\% flagged. The Y-axes are limited to 10\% for better visualization. The region enclosed by the black dotted circle is $z\sim0.32$ which shows $>$10\% flagging for most pointings. The shaded region is for $z\sim0.44$ which shows $\lesssim$10\% flagging for most pointings, thus used for further analysis. Panels are arranged the same as Figure \ref{flags_chan}.}
\label{flags_zoom}
\end{figure*}

Flagging percentages can also vary over baselines for the same observation block. Since a very large amount of flagged data may produce spurious features in the final power spectrum, we filter out baselines having more than 20\% flagged data for our analyses. Red points in Figure \ref{flags_stem} show the flagging percentage for each pointing in the dashed sub-band (i.e. centred at $\sim$986\,MHz) using only baselines with $\geq$80\% unflagged data. The flagging percentage as a function of channels per pointing is plotted in Figure \ref{flags_zoom} for the \texttt{LOW} band. The Y-axes are zoomed-in to 10\% for better visualisation. The sub-band corresponding to z$\sim$0.32 (1054-1100\,MHz) is shown by dotted boundaries. It is seen that while excluding heavily flagged baselines does improve the overall behaviour, this sub-band still has a few channels with flagging fractions higher than 10\%, specifically towards the higher frequency end. Conversely, the shaded region in Figure \ref{flags_zoom} (with frequency 962.55-1008.42\,MHz corresponding to z$\sim$0.44) shows that imposing the baseline cut makes the sub-band cleaner while maintaining an almost uniform flag level throughout for almost all the pointings. Thus, we confine our subsequent analysis for this work to $z\sim0.44$ sub-band i.e. the shaded regions in Figure \ref{flags_zoom}, which is consistent with one of the sub-bands used in \citet{paul2023detection}. Unlike \citet{paul2023detection}, we have not performed any in-painting for the flagged data, since, as mentioned, the flagging percentages are almost uniform across the sub-band. Future work will investigate other frequency windows and how the power spectrum is affected by incorporating visibilities with higher flagging levels and data in-painting.

\subsection{Thermal Noise Simulations}
\label{noise_sim}

This section describes the methodology to simulate the noise of our observations. The noise estimation is important because the power spectrum estimator described in Section \ref{estimator} uses inverse noise variance weighting. Since astrophysical foregrounds in our scales of interest are expected to have negligible circular polarization, we can use Stokes V data to estimate the thermal noise amplitude. It was shown in \citet{sourabh_2021, paul2023detection} that the MeerKAT data has thermal noise consistent with the Stokes V. Gridding the Stokes V data similar to the total intensity data as described in Section \ref{uv-grids}, the noise amplitude for $N$ baselines in a uv-cell with average visibility $V$ is  $\sigma_N = \textrm{std}(V\sqrt{N})$. 

To estimate the thermal noise per pointing, we simulate the noise power spectrum where we replace the gridded visibilities in each uv-cell (as described in Section \ref{uv-grids}) with a random realisation of a zero mean complex circular Gaussian with a standard deviation of:
\begin{equation}
\sigma_{N} = \frac{2k_B\mathrm{T_{sys}}}{A_e \sqrt{\delta\nu\delta t}}
    \label{noise_eq}
\end{equation}

where k$_B$ is the Boltzmann constant, the channel width $\delta\nu$ is 104.5\,kHz and time resolution is $\delta t$=8\,s. Since we are generating the noise from the gridded visibilities, the generated noise realizations are divided by the square root of the counts per cell. The natural sensitivity of the instrument is $A_e/\mathrm{T_{sys}}$ with system temperature $\mathrm{T_{sys}}$ and effective area $A_e$ having anticipated values of 6.22\,m$^\mathrm{2}$/K for MeerKAT \citep{Specs}. The expected noise is also calculated from Stokes V as described above, and we obtain a value of $\sim$6.49 \,m$^\mathrm{2}$/K. 

\begin{figure}
\includegraphics[width=\columnwidth]{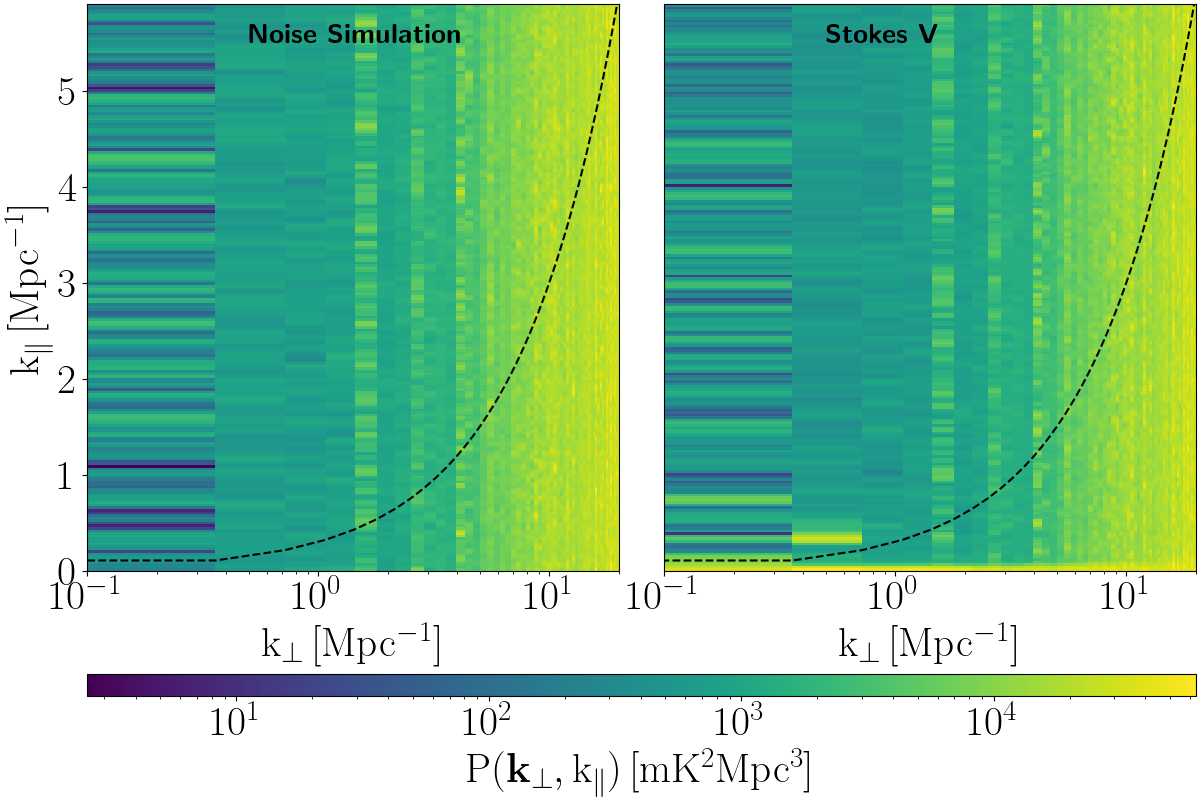}
\includegraphics[width=\columnwidth, trim={0.1cm 0.3cm 0.1cm 0.2cm},clip]{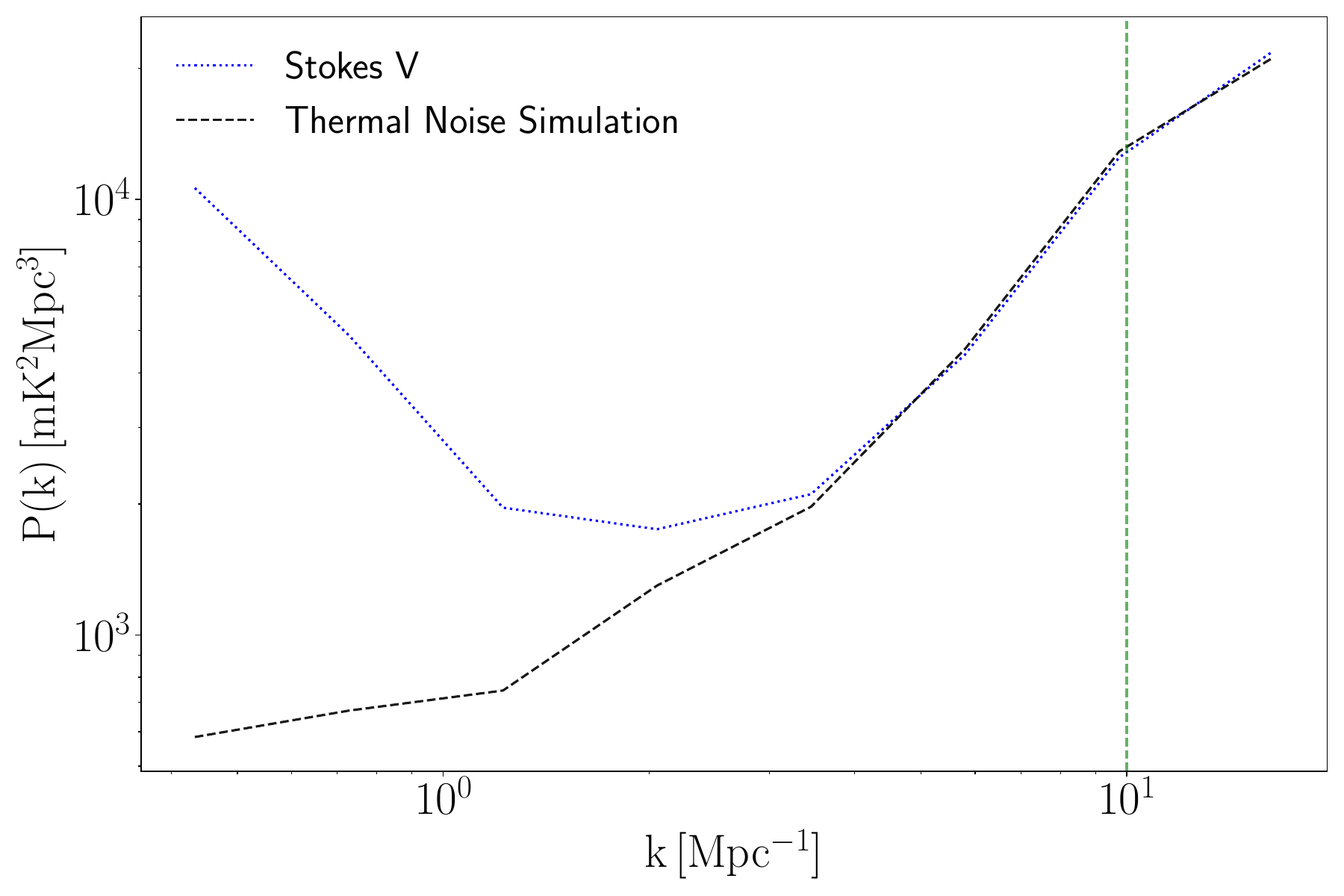}
\caption{\textit{\textbf{(top)}} Cylindrical power spectrum for a single realization of thermal noise simulation (left panel) and gridded Stokes V visibilities (right). \textit{\textbf{(bottom)}} The spherically averaged power spectrum of the thermal noise simulation compared against Stokes V. The green dashed curve is the $k$ cut-off used for final power spectrum measurements. The simulated noise is consistent with Stokes V, except in the lower k$_\perp$-k$_\parallel$ plane, as evident from both panels.}
\label{tn_power}
\end{figure}

We simulate 10$^5$ noise realizations for each pointing and the average value for each $k$ pixel gives the noise variance. Figure \ref{tn_power} shows the Stokes V and noise power spectra for COSMOS\_2  with the cylindrical power on the top panel and the spherical power on the bottom panel. The top-left panel demonstrates the noise simulations, top-right shows the Stokes V power on the right. The black dashed curve on the bottom panel shows the spherical power spectrum of the noise simulations while the blue dotted curve shows the same for Stokes V. From the top panel, the Stokes V power closely resembles the simulated noise power except at the lowest region on the  $k_{\parallel}$ - k$_\perp$ plane. This is also reflected in the bottom panel of Figure \ref{tn_power}, where the simulated spherical noise power deviates from Stokes V at small scales. The value of $k$ for which Stokes V deviates from simulated thermal noise estimates is similar to that seen in \citealt{paul2023detection}. Although Figure \ref{tn_power} shows a particular pointing, the same trend is seen for all the pointings of COSMOS data. 

The sparser sampling of long baselines (or large $k_{\perp}$ values) compared to the short ones leads to higher noise in the former case. However, the line-of-sight i.e. $k_\parallel$ direction does not have this non-uniformity. Thus, the noise is expected to vary more in the $k_{\perp}$ direction, which predominantly contributes to the noise power spectrum. This leads to the increase in both the noise power and Stokes V powers seen at large $k$ values. The green dashed line at  $k\sim$10\,Mpc$^{-1}$ corresponds to $k_{\perp}$ beyond which baselines are sparsely sampled. Any measurement beyond these values would be noise-like, demonstrating the need for a cut-off value for the highest usable $k_{\perp}$ value for our final power spectrum measurements. 

The deviation between the simulated noise and Stokes V is due to power leakage into Stokes V. The structure is clearly visible at low $k_{\perp}$, as seen in both Figure \ref{tn_power} and \ref{stokesV_ratio}. These modes are dominated by contributions from the central part of the $uv$ plane, resulting in high signal-to-noise and more evident systematic features. Conversely, the large k values being noise-dominated, we do not see this feature so clearly.

Figure \ref{stokesV_ratio} shows the ratio of the Stokes V power over the noise power for all MIGHTEE-COSMOS fields, all of which show the excess at the smallest $k_\parallel$ across all $k_\perp$. However, we use only the longer baselines (where the noise simulations and Stokes V match quite well) to normalise the simulated noise power. Thus, this is not a cause of concern for our results.

\subsection{Power Spectrum Analysis}
In this section, we present tests performed in the power spectrum domain. We start with the 3D power spectrum (Equation \ref{2dps_eq}) obtained from gridded continuum-subtracted visibility cubes as described in Section \ref{uv-grids}. The power spectrum is multiplied by a Blackman-Harris window, which suppresses the foreground leakage at higher $k_{\parallel}$ modes. We use both auto and cross-power spectra to test for residual systematic contributions.

\subsubsection{Auto Power Spectrum per Pointing}
\label{cyl_p}
\begin{figure*}
\centering
\begin{subfigure}{0.3\textwidth}
    \includegraphics[width=\textwidth]{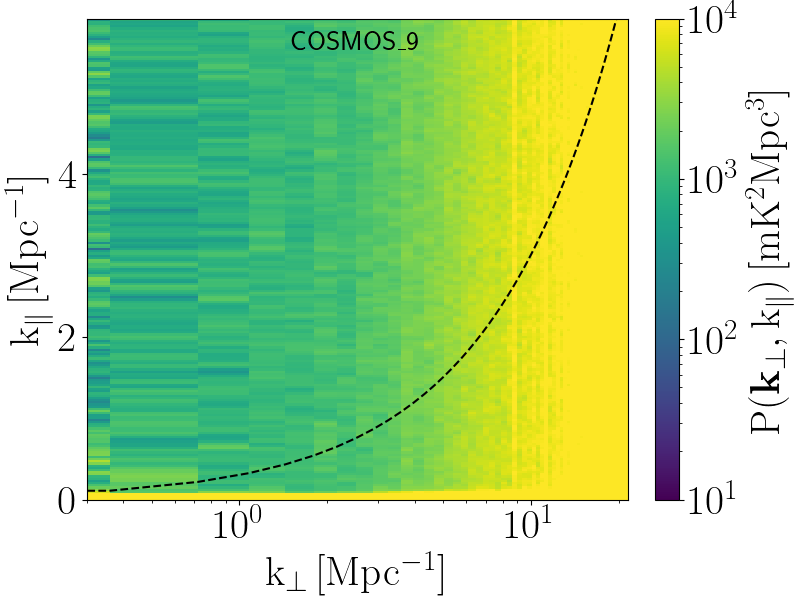}
    \label{fig:cosmos9}
\end{subfigure}
\hspace{1pt}
\begin{subfigure}{0.3\textwidth}
    \includegraphics[width=\textwidth]{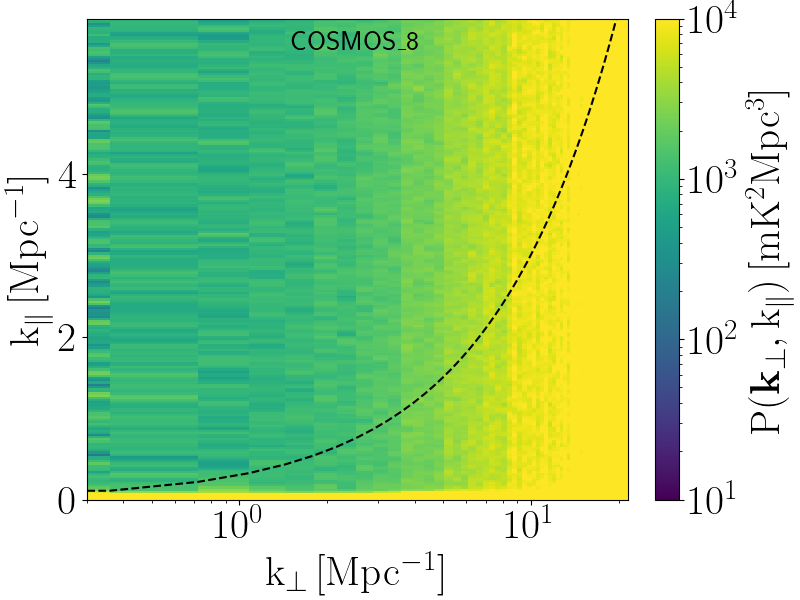}
    \label{fig:cosmos8}
\end{subfigure}
\hspace{1pt}
\begin{subfigure}{0.3\textwidth}
    \includegraphics[width=\textwidth]{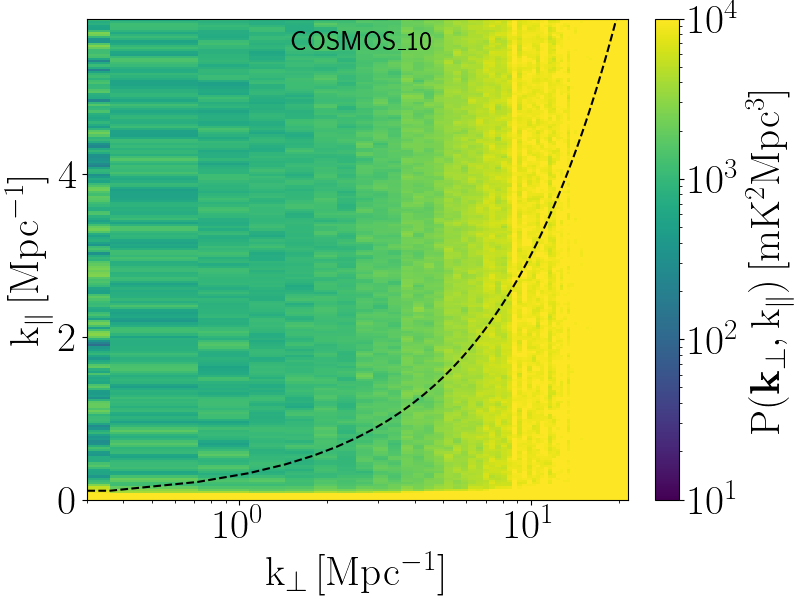}
    \label{fig:cosmos10}
\end{subfigure}
\begin{subfigure}{0.3\textwidth}
    \includegraphics[width=\textwidth]{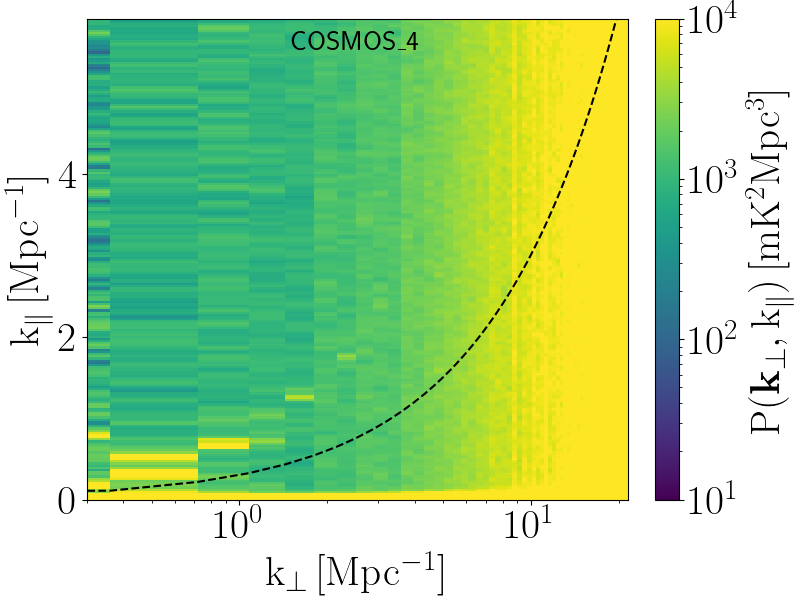}
    \label{fig:cosmos4}
\end{subfigure}
\hspace{1pt}
\begin{subfigure}{0.3\textwidth}
    \includegraphics[width=\textwidth]{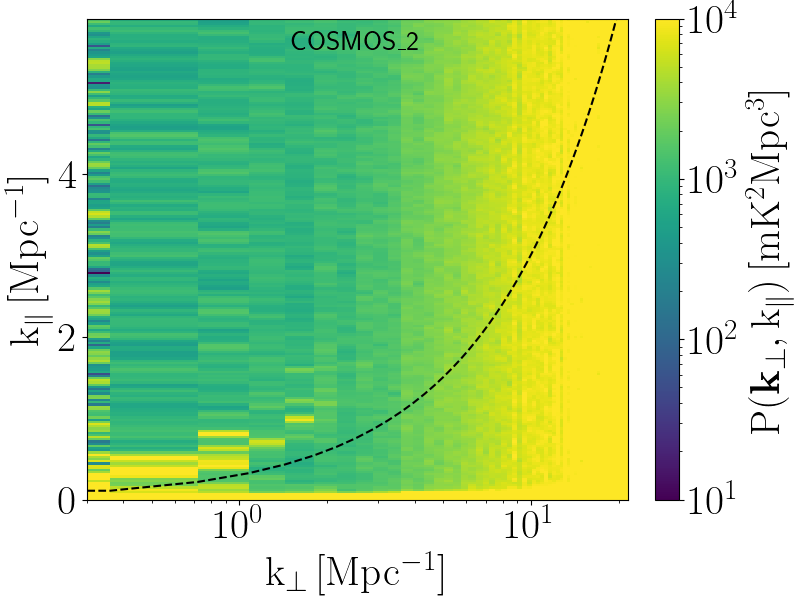}
    \label{fig:cosmos2}
\end{subfigure}
\hspace{1pt}
\begin{subfigure}{0.3\textwidth}
    \includegraphics[width=\textwidth]{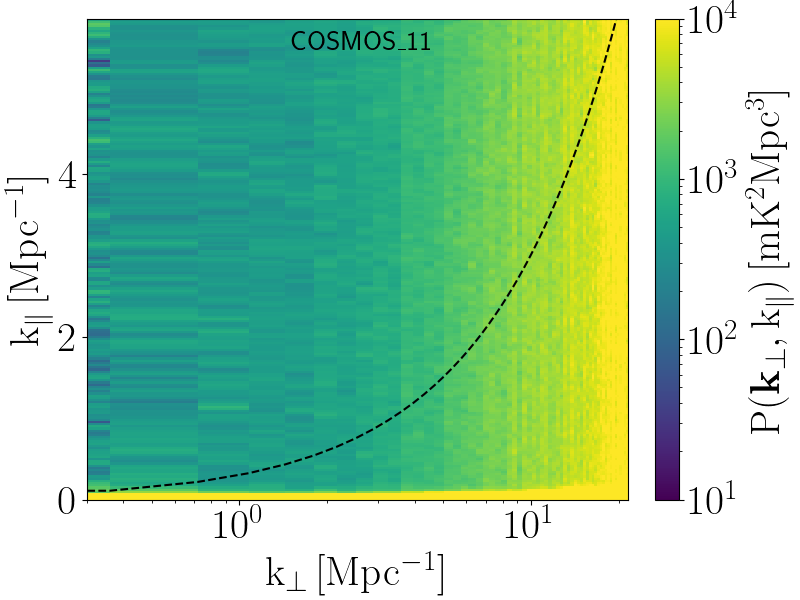}
    \label{fig:cosmos11}
\end{subfigure}
\begin{subfigure}{0.3\textwidth}
    \includegraphics[width=\textwidth]{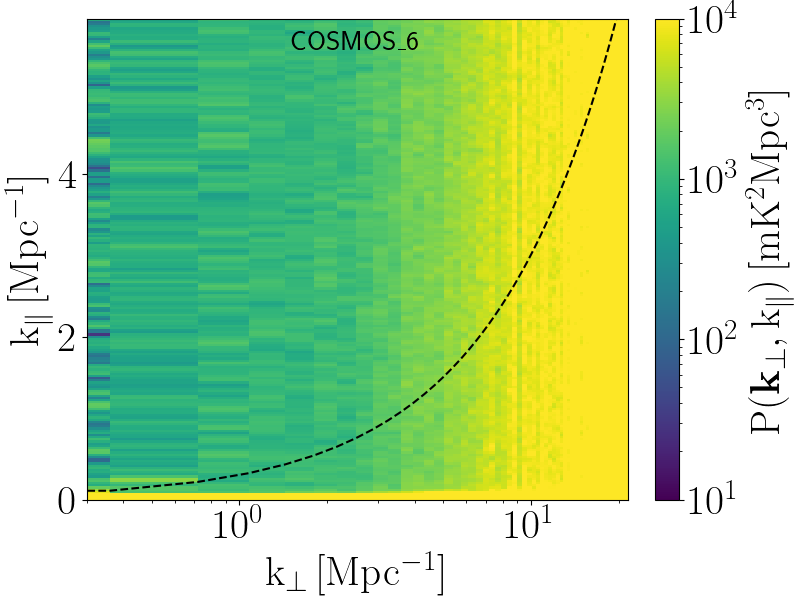}
    \label{fig:cosmos6}
\end{subfigure}
\hspace{1pt}
\begin{subfigure}{0.3\textwidth}
    \includegraphics[width=\textwidth]{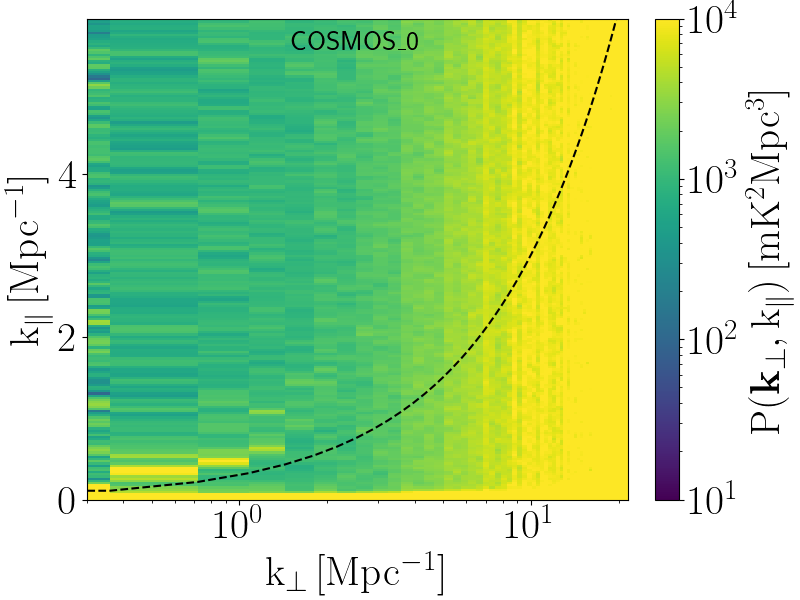}
    \label{fig:cosmos}
\end{subfigure}
\begin{subfigure}{0.3\textwidth}
    \includegraphics[width=\textwidth]{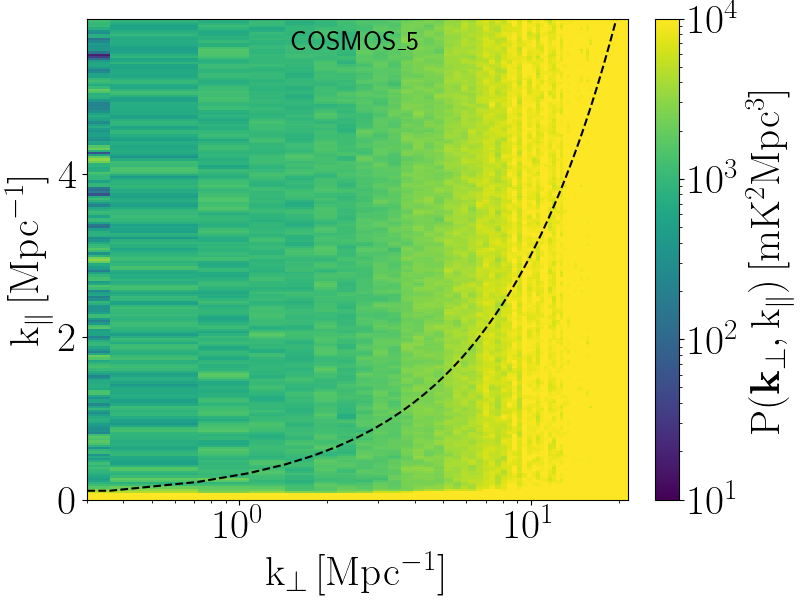}
    \label{fig:cosmos5}
\end{subfigure}
\begin{subfigure}{0.3\textwidth}
    \includegraphics[width=\textwidth]{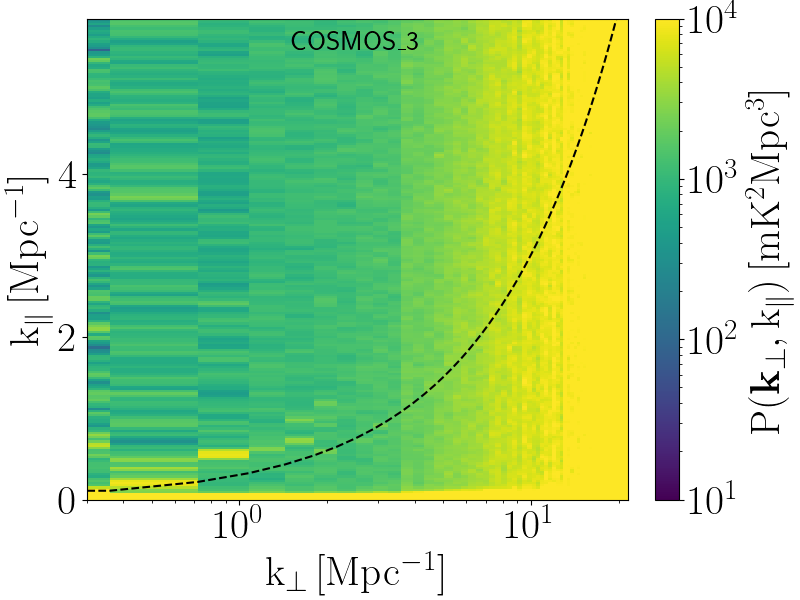}
    \label{fig:cosmos3}
\end{subfigure}
\begin{subfigure}{0.3\textwidth}
    \includegraphics[width=\textwidth]{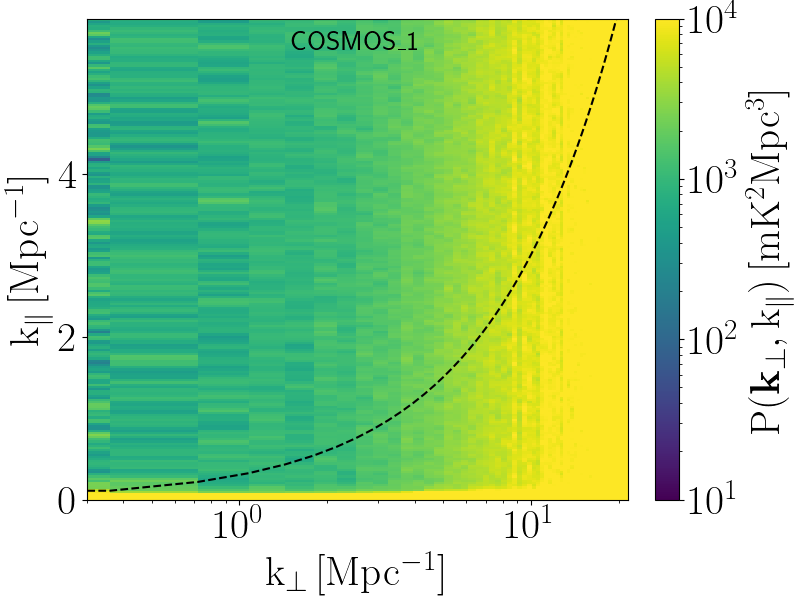}
    \label{fig:cosmos1}
\end{subfigure}
\hspace{1pt}
\begin{subfigure}{0.3\textwidth}
    \includegraphics[width=\textwidth]{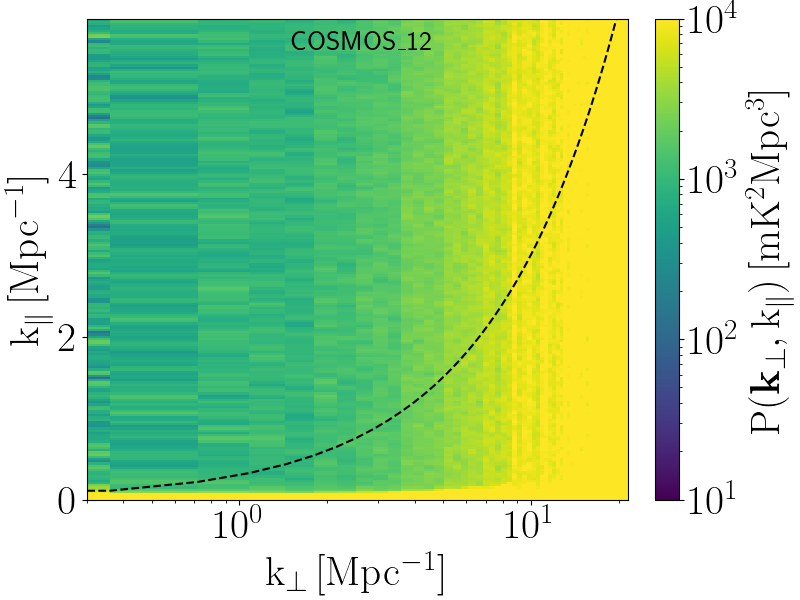}
    \label{fig:cosmos12}
\end{subfigure}
\begin{subfigure}{0.3\textwidth}
    \includegraphics[width=\textwidth]{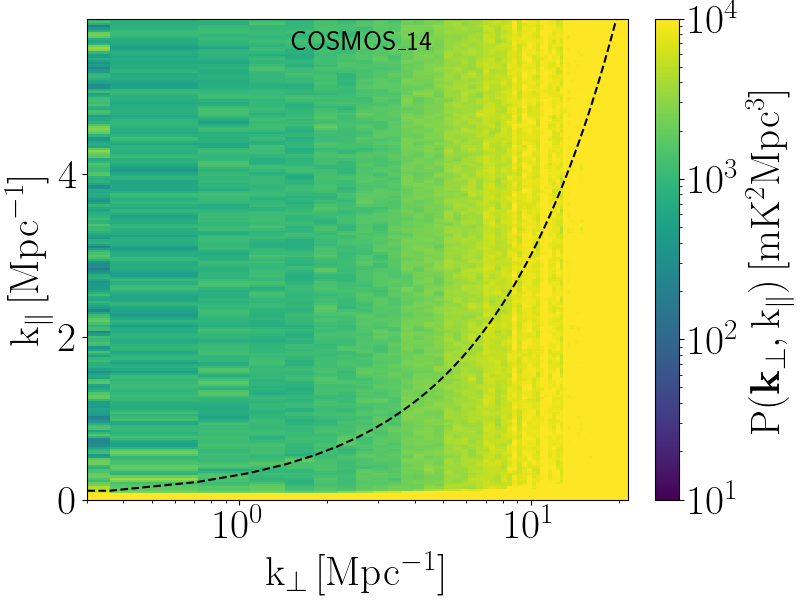}
    \label{fig:cosmos14}
\end{subfigure}
\hspace{1pt}
\begin{subfigure}{0.3\textwidth}
    \includegraphics[width=\textwidth]{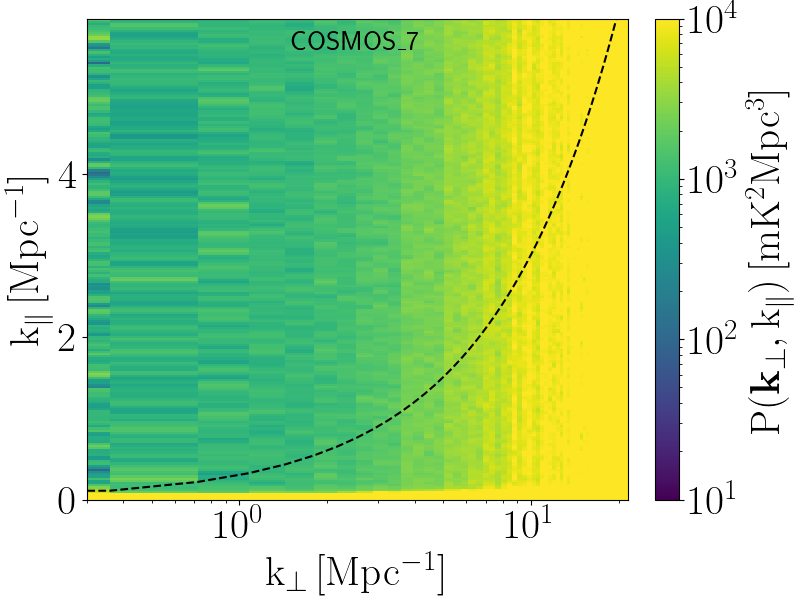}
    \label{fig:cosmos7}
\end{subfigure}
\begin{subfigure}{0.3\textwidth}
    \includegraphics[width=\textwidth]{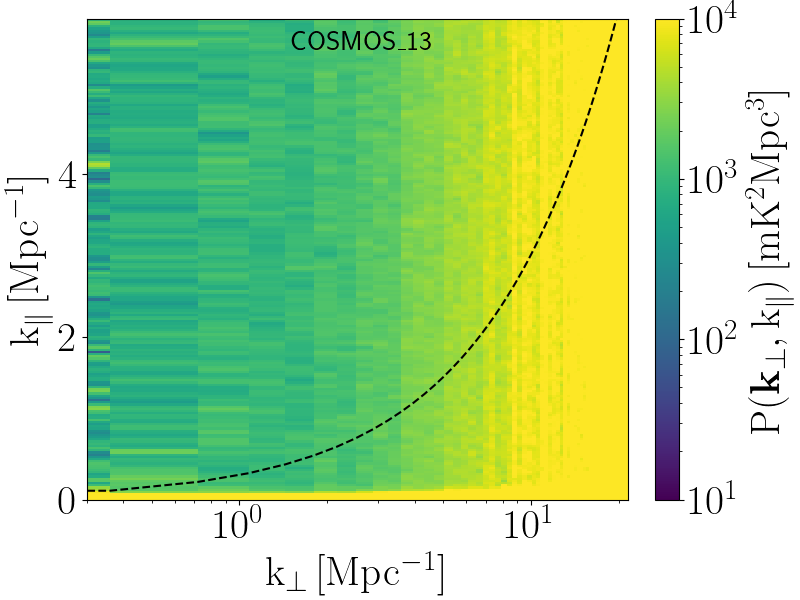}
    \label{fig:cosmos13}
\end{subfigure}
\caption{Cylindrical averaged auto-correlation power spectrum of the continuum-subtracted MIGHTEE-COSMOS data. Each panel is the auto-correlation power spectrum of a single pointing, as indicated in the panel name. Each row represents pointings at the same declination. There are visible signatures of systematics present in some pointings, near the blacked dashed curve (${k}_{\parallel} \sim 0.3 {k}_{\perp}$, horizon line) for small ${k}_{\perp}$ values. }
\label{2dps_point}
\end{figure*}

In this section, we describe tests done with the auto-power spectrum (i.e. $j=i$ in \ref{2dps_eq}) for each pointing. The presence of residual systematics that do not correlate out is easier to detect in the auto-power spectrum. Thus, we use this test to find the amount and location of contamination expected in the final averaged power. The cylindrical auto-power spectrum for each pointing is shown in Figure \ref{2dps_point}. As done previously, the pointings are named according to \autoref{table:pointings}, and the same declinations are grouped in rows. From the individual panels of Figure \ref{2dps_point}, clear systematic contamination is seen inside the observation window i.e. above the horizon line for some pointings.  The pointings COSMOS\_2, COSMOS\_4, COSMOS\_0 and COSMOS\_3. COSMOS\_2, COSMOS\_3 and COSMOS\_4, which show higher levels of flagging, also show excess power beyond the horizon limit. These results indicate that these pointings are affected more by systematics. We show further evidence of contaminations in Figure \ref{stokesV_ratio}, where these pointings have excess power in Stokes V at small ${k}_{\perp}$ modes. The first few channels of COSMOS\_0 have a high flagging percentage in the sub-band used (first panel in the third row of Figure \ref{flags_zoom}). COSMOS\_0 also exhibits systematic signatures in the small $k_{\perp}$ modes beyond the horizon line, as evident from Figure \ref{2dps_point} (and from Figure \ref{stokesV_ratio}). The exact reason for the observed contamination is unknown. One possibility is that the observing tracks have the earliest starting times around 15:30 local time (or 13:30 UTC) and also cover local sunset, which affects the data. Figure \ref{2dps_point} does not point to any clear declination dependence of systematics.

\begin{figure*}
\centering
\includegraphics[width=\columnwidth, trim={0.15cm 0.1cm 0.2cm 0.1cm},clip]{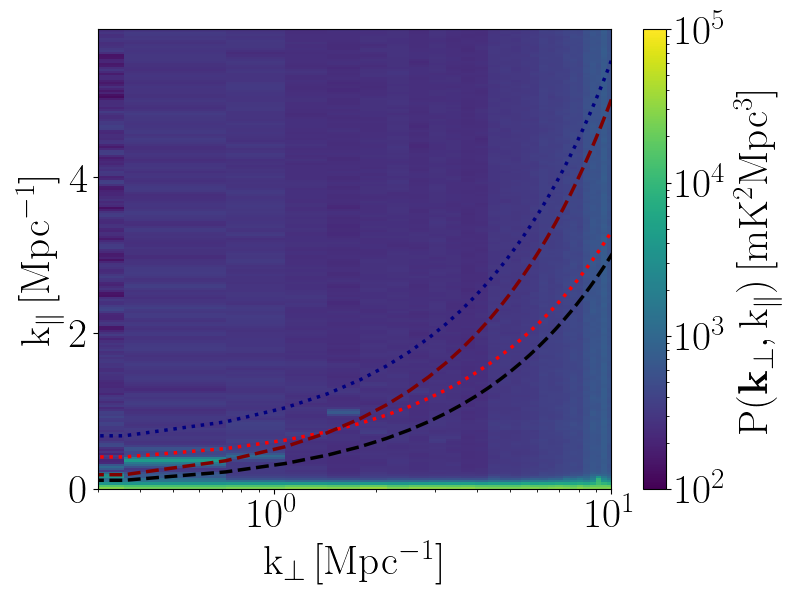}
\includegraphics[width=\columnwidth]{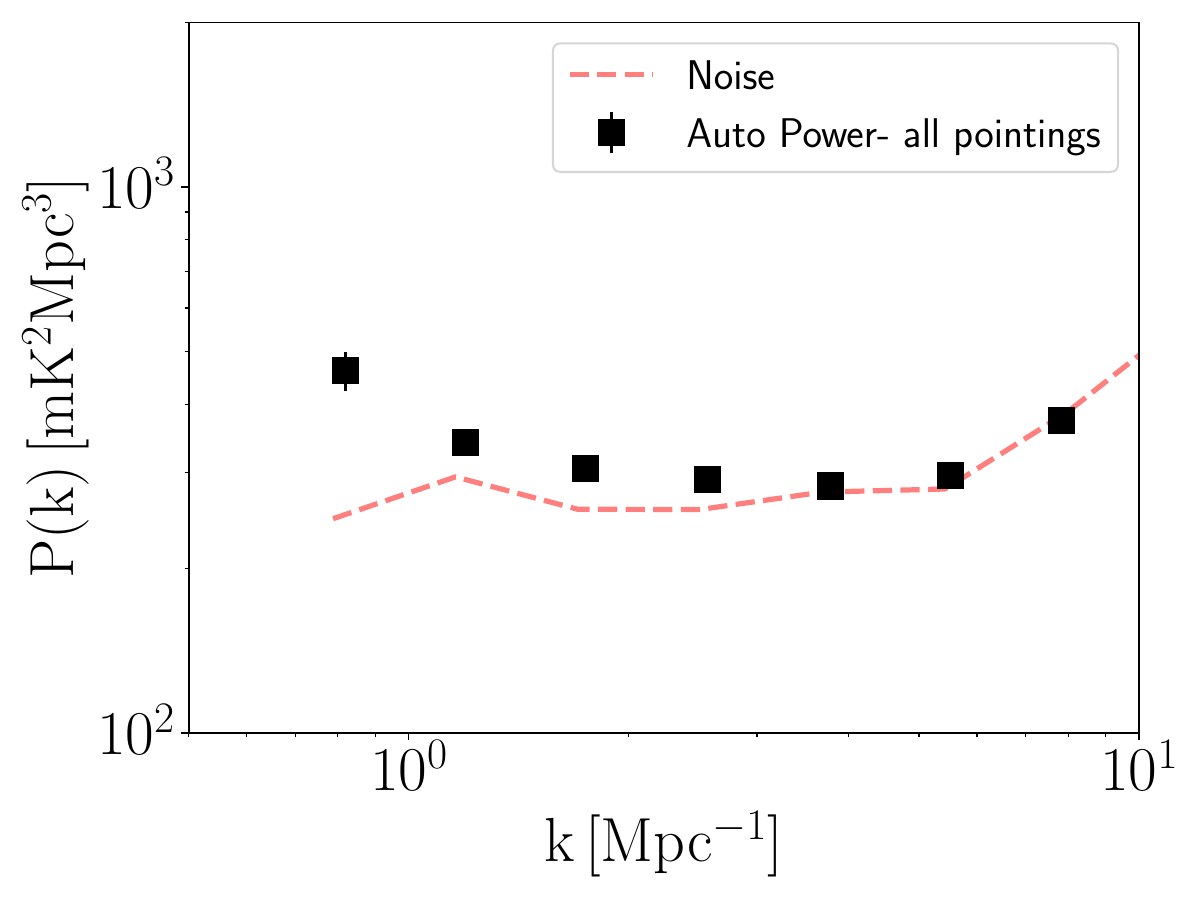}
\caption{(\textit{\textbf{left}}) Cylindrical auto-power spectrum from continuum subtracted visibilities for all pointings combined. The dashed and dotted curves are different foreground cuts to avoid systematics affected regions: horizon limit i.e. ${k}_{\parallel} \sim 0.3 {k}_{\perp}$ (black dashed curve), ${k}_{\parallel} \sim 0.5 {k}_{\perp}$ (maroon dashed curve), ${k}_{\parallel} \sim 0.3{k}_{\perp}+0.3$ (red dotted curve) and ${k}_{\parallel} \sim 0.5 {k}_{\perp}+0.5$ (blue dotted curve). (\textit{\textbf{right}}) Spherically averaged auto-power spectrum of continuum-subtracted MIGHTEE-COSMOS data. The black squares show the power spectrum calculated from the data using modes beyond the horizon limit, i.e. ${k}_{\parallel} \sim 0.3 {k}_{\perp}$}, and the red dashed line shows the average thermal noise power from all pointings.
\label{2dps_auto}
\end{figure*}

As a final test, we show the averaged auto-power spectrum of the combined MIGHTEE-COSMOS pointings in Figure \ref{2dps_auto}. The left panel shows the cylindrical power, while the right panel shows the spherically averaged power spectrum. In the left panel, the excess power seen beyond the horizon limit (black dashed curve) $k_\perp \,\lesssim$1\,Mpc$^{-1}$ originates from features seen for some of the cases in Figure \ref{2dps_point}. The smallest $k_\perp$-bin ($<$0.5\,Mpc$^{-1}$) is noise-dominated and may have subdominant systematics. Insufficient signal-to-noise prevents us from properly determining the major contaminant in this bin. We show the spherical auto-power spectrum marked by black squares in the right panel of Figure \ref{2dps_auto}. We use the modes outside the horizon limit of ${k}_{\parallel} \sim 0.3 {k}_{\perp}$  (i.e. marked by the black dashed curve on the left panel). The average thermal noise is shown by the red dashed line. The auto-power measurements follow the noise except the smallest $k$-bin, where the dominant contribution comes from the $k$-modes contaminated by systematics. Removing the noise bias should give us a better estimation of the residual power in the data, which is discussed later.

\subsubsection{Effect of Changing Measurement Window}
\label{fg_cut_sec}
\begin{figure}
\centering
\includegraphics[width=\columnwidth, height=6.5cm]{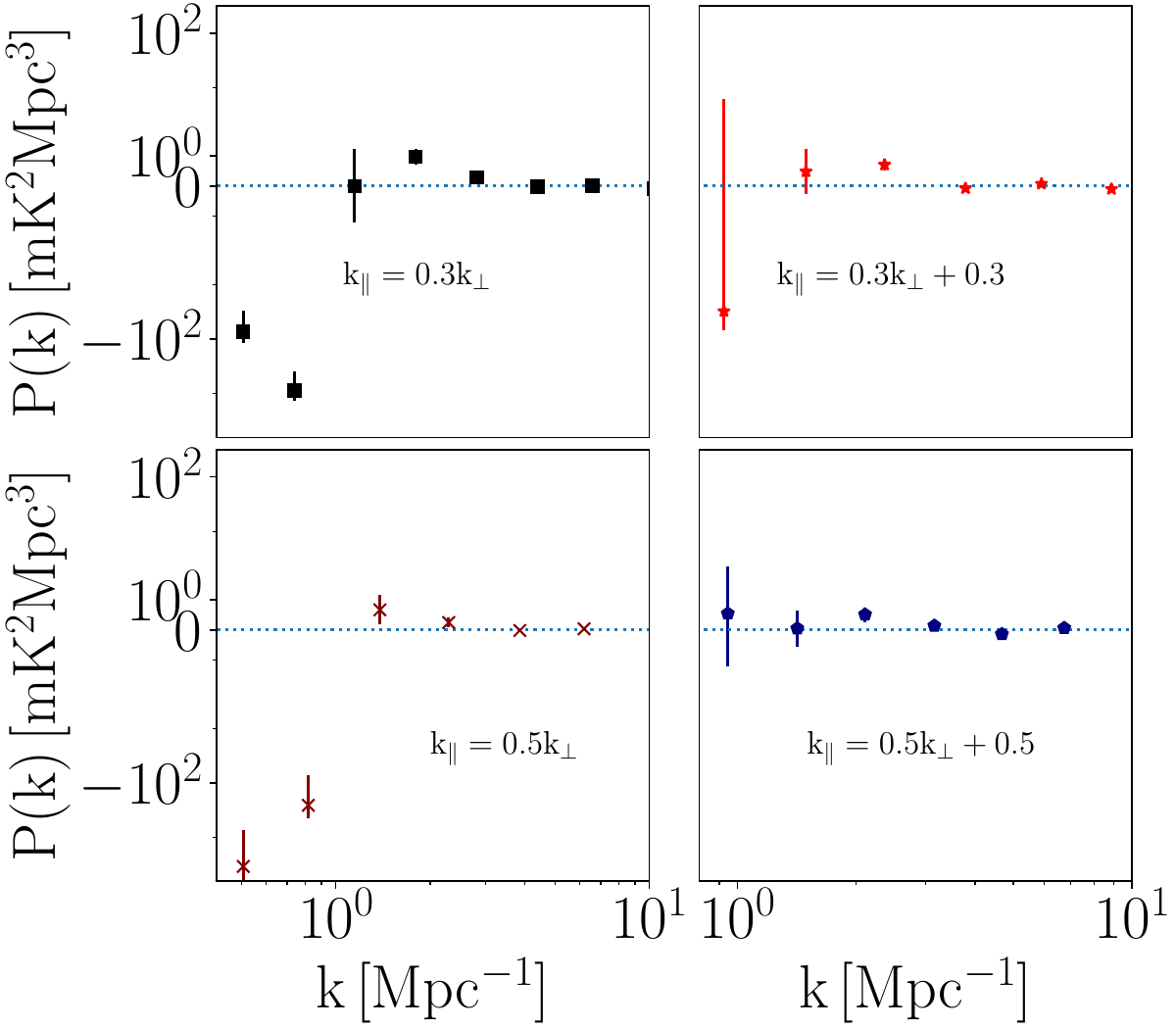}
\caption{Spherical cross-power spectrum with different foreground cuts shown in the left panel of Figure \ref{2dps_auto}- horizon limit of ${k}_{\parallel} \sim 0.3 {k}_{\perp}$ (black squares, upper left panel), ${k}_{\parallel} \sim 0.3{k}_{\perp}+0.3$ (red stars, upper right panel), ${k}_{\parallel} \sim 0.5 {k}_{\perp}$ (maroon crosses, lower left panel) and ${k}_{\parallel} \sim 0.5 {k}_{\perp}+0.5$ (blue pentagons, lower right panel).}
\label{fg_cuts}
\end{figure}

We test whether excluding the contaminated modes beyond the horizon limit significantly improves the 1D power spectrum. We test three different foreground avoidance criteria (marked by different coloured curves in \autoref{2dps_auto}):  ${k}_{\parallel} \sim 0.5 {k}_{\perp}$ (zone beyond the magenta dashed curve), ${k}_{\parallel} \sim 0.3 {k}_{\perp} +0.3$ (zone beyond red dotted curve) and ${k}_{\parallel} \sim 0.5 {k}_{\perp}+0.5$ (zone beyond the blue dotted curve). The foreground cuts are used to measure the cross-power spectrum (to remove the effect of noise bias) and compare with the horizon limit cuts. Figure \ref{fg_cuts} shows the spherical cross-power spectra in different subplots- the top left is the spherical power with the original horizon limit, the top right is ${k}_{\parallel} \sim 0.3{k}_{\perp}+0.3$, bottom left is ${k}_{\parallel} \sim 0.5 {k}_{\perp}$ and bottom right is  ${k}_{\parallel} \sim 0.5 {k}_{\perp}+0.5$. It is seen from the bottom right sub-plot that while there is an improvement in the smallest $k$ bin, the overall power spectrum is still relatively noisy. The other two cases show no marked improvement compared to the horizon cut. The bottom right panel ($\sim 0.5 {k}_{\perp}+0.5$) does show some improvement in the first bin, but it also excludes a large number of modes. Since more stringent cuts lower the signal-to-noise, with improvements only in the smallest $k$ bin, we use the original measurement window beyond the instrument horizon for our final measurements. It should also be noted from Figure \ref{2dps_auto} that most modes beyond the foreground wedge region are clean of systematics and a measurement window, excluding the few contaminated modes at small $k_\perp$, but including everything beyond the wedge boundary can enhance the signal-to-noise. However, such measurement windows would be specific to this particular data set, and therefore, we use ${k}_{\parallel} \sim 0.3 {k}_{\perp}$ limit to keep the analysis general.

\subsubsection{Power Distribution per $k$ bin}
\label{hists}
\begin{figure*}
\centering
\includegraphics[width=0.8\textwidth]{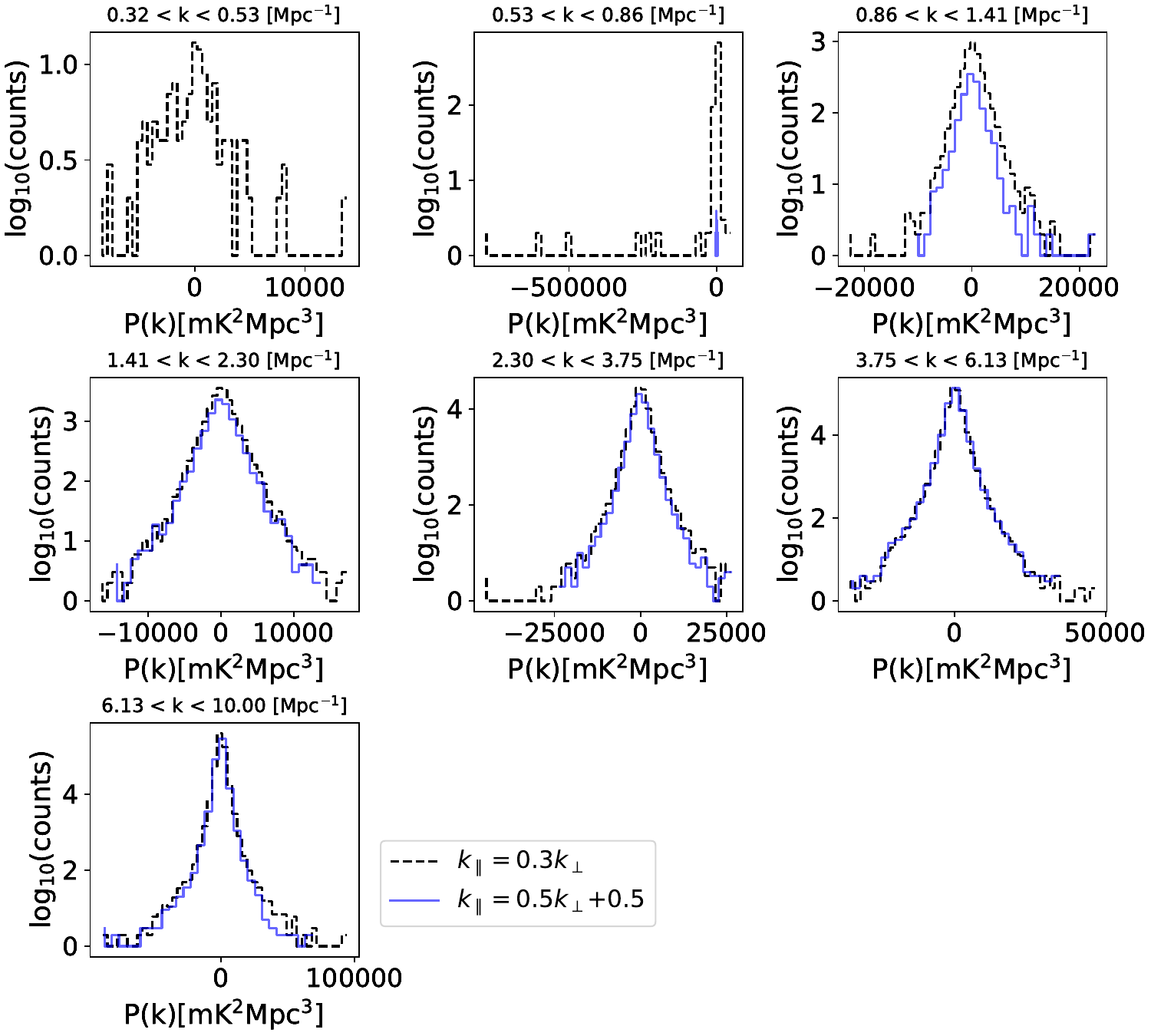}
\caption{Distribution of the power spectrum per $k$ bin from cross-correlating the odd and even scans for two measurement windows. The black dashed histograms are for the ${k}_{\parallel} \sim 0.3 {k}_{\perp}$ cut and blue solid histogram is for ${k}_{\parallel} \sim 0.5 {k}_{\perp}+0.5$ cut. The latter cut excludes the first $k$ bin, hence it contains only the black curve. The latter cut excludes more modes resulting in difference in the histogram count. Using the more stringent foreground cuts results in an improvement in the smallest $k$ mode, but there is no significant difference seen for the other modes.}
\label{power_hist}
\end{figure*}

In this section, we examine the statistics of the averaged cross-power spectrum. The Fourier modes are binned into logarithmic $k$ bins, and the resulting distribution in each $k$ mode is plotted in Figure \ref{power_hist}. Two cases are shown with different measurement windows - ${k}_{\parallel} \sim 0.3 {k}_{\perp}$ (black dashed histograms) and ${k}_{\parallel} \sim 0.5 {k}_{\perp}+0.5$ (blue solid histogram). The level of outliers in each $k$-bin can be detected from the distribution of power in it. The noise is expected to be Gaussian random, thus heavily skewed distributions indicate the power to be systematics-dominated rather than noise-dominated. The ${k}_{\parallel} \sim 0.5 {k}_{\perp}+0.5$ cut does not measure the smallest bin, hence the first panel only contains the black histogram.

The smallest scale that can be measured with both cuts is $\sim$0.7\,Mpc$^{-1}$. The blue histogram in the corresponding bin (second panel in the first row of Figure \ref{power_hist}) has a very small spread. Hence, using the stringent cut improves the measurement in the smallest k-mode by both (also evident from the bottom right panel of Figure \ref{fg_cuts}). However, the histograms in other bins are both similarly distributed. Thus, the stricter cut does not show a significant improvement over using the horizon limit. Additionally, the smaller measurement window also reduces the number of accessible modes and the counts in each bin. The skewed histogram in the first panel also shows that a large portion of the discernible contamination is from the systematics in the shortest baselines contributing dominantly in this bin. Nevertheless, Figure \ref{power_hist} reinforces that using a more stringent cut does not improve the measurements considerably. Additionally, the limit ${k}_{\parallel} \sim 0.3 {k}_{\perp}$ gives a larger number of $k$ modes and sufficiently large counts across them, hence it is used for our final measurements. 

\subsection{Dropout Tests with Different Pointings} 
\label{jackknife}

As evident from Section \ref{cyl_p}, some pointings have systematics in the small $k_{\perp}$ modes.  Hence, it is plausible that removing these contaminated pointings can improve the overall results, with the obvious caveat that the large amount of data removed will lower the signal-to-noise. Since the MIGHTEE-COSMOS tracks the target continuously for about an hour before the intermediate secondary calibrator scan, time-dependent systematics present in one or more scans may impact the overall results. However, if these contaminations are transient (i.e. present in one scan but absent in the others), they would drop out on correlating different scans. But if present over larger time scales, the effects may be enhanced (or may not correlate out as effectively) on cross-correlating. Thus, we do drop-out tests by removing each pointing entirely to check for time-dependent systematics in the data.

\begin{figure}
\centering
\includegraphics[width=0.5\textwidth]{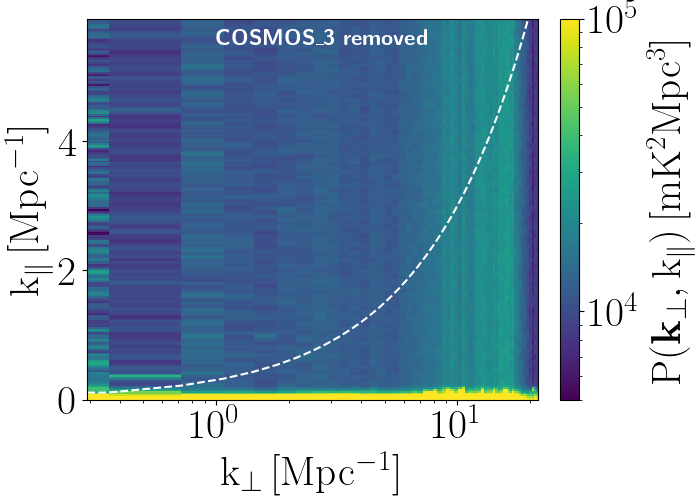}
\includegraphics[width=0.5\textwidth]{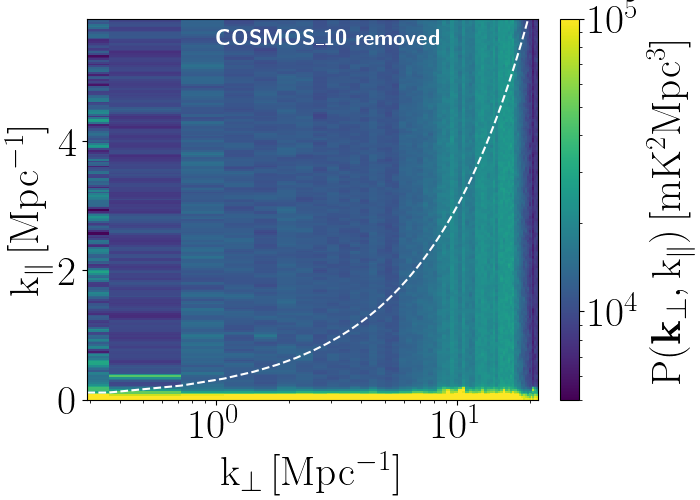}
\caption{Cylindrical auto-power spectrum for dropout tests. \textit{(\textbf{top panel})} This shows the result on removal of COSMOS\_3, a pointing with clear systematic contamination seen in Figure \ref{2dps_point}. While some of the bins show lower levels of excess power, there are still some persistent features due to contaminants in other pointings.  \textit{(\textbf{bottom panel})} This shows the case on removing COSMOS\_10, one of the clean pointings. As expected, there is no clear improvement seen on removing a pointing with no visible contamination. Both cases use continuum-subtracted visibilities.}
\label{cyl_jack}
\end{figure}

In Figure \ref{cyl_jack}, we show the averaged cylindrical power spectra after dropping out two pointings - COSMOS\_3 and COSMOS\_10. They represent pointings for which Figure \ref{2dps_point} shows clear systematics in the measurement window (upper panel) and a case with no visible contamination (lower panel). We see that combining pointings makes the level of contamination average down, but only very slightly (for instance as seen in the top panel). The ``cleanest" results would of course include only clean pointings, but given the volume of data that we would lose, it is not feasible for this work.
 
\begin{figure}
\centering
\includegraphics[width=\columnwidth,trim={0.3cm 0.3cm 0.3cm 0.3cm},clip]{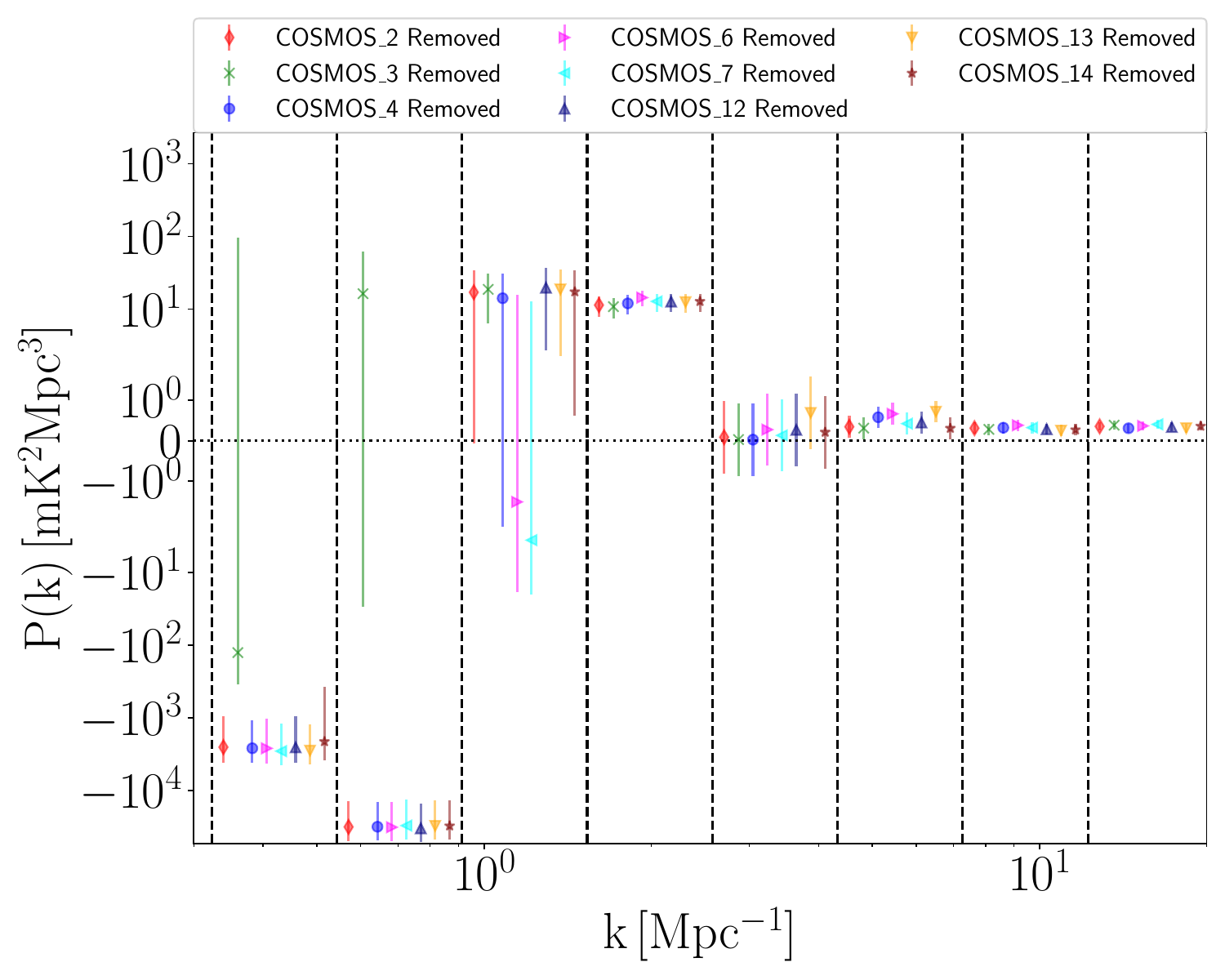}
\caption{Spherically averaged cross-power spectrum for the drop-out tests with individual pointings removed. The samples shown are- COSMOS\_2 (red diamonds), COSMOS\_3 (green crosses), COSMOS\_4 (blue circles), COSMOS\_10 (navy triangles), COSMOS\_6 (magenta right carat), COSMOS\_7 (cyan left carats), COSMOS\_12 (orange inverted triangles) and COSMOS\_14 (maroon stars). The error bars are the 1$\sigma$ uncertainties for each case.}
\label{sphe_jack}
\end{figure}

Figure \ref{sphe_jack} shows the 1D cross-power spectrum for the drop-out tests demonstrated in Figure \ref{cyl_jack}. Similar to Section \ref{fg_cut_sec}, we use cross-power to remove the noise bias. We use 7 logarithmic bins for all the cases, but Figure \ref{sphe_jack} has the $k$ values slightly displaced for the different samples for better visualization. It is seen that the first $k$ mode (centred at 0.34\,Mpc$^{-1}$) is negatively biased, indicative of some systematics in the smallest k$_\perp$ mode. The second bin centred at 0.57\,Mpc$^{-1}$ shows improvement on removing COSMOS\_3 (though the error bar is still large). For the other bins, the results are sufficiently consistent across all the cases. Investigations into COSMOS\_3 show that while most other pointings start observing late afternoon and get only the local sunset, it starts observation near local noon. The observation time may be the cause of the slightly greater amount of contamination in the data. It is also possible that local ionospheric conditions on this particular day were more turbulent. It also shows that subdominant systematics can be a problem with the method used since it is not always obvious over the noise. However, it is expected that with the addition of more data, we can apply more stringent cuts without compromising too much on the signal-to-noise and thus improve the results.

\subsection{Effect of Continuum Subtraction}
\label{signal loss}
The visibility domain continuum subtraction described in Section \ref{reduction} could potentially cause \hi signal loss by removing \hi along with the continuum emission. In Figure \ref{2dps_auto}, the amplitude of the foreground wedge is an order of magnitude lower than expected. The continuum emission from foregrounds, the dominant contributor to the wedge power, is removed from the visibilities used here for \hi intensity mapping. 

The visibility domain subtraction uses the model created from the deep MIGHTEE continuum images and subtracts the continuum model from self-calibrated data. The obtained residuals are used for our analysis, so we calculate the ratio of this continuum-subtracted visibility with the visibility containing the continuum (or foregrounds). If the process does not remove any significant amount of \hi, the ratio in the window should be $\sim$1, while a significantly smaller value will be seen in the wedge. We also measured the cylindrical power spectrum of the subtracted continuum model. In the absence of \hi in the model, we would expect negligible power in the region outside the wedge. 
 
\begin{figure*}
\centering
\includegraphics[width=\columnwidth, height=2.3in]{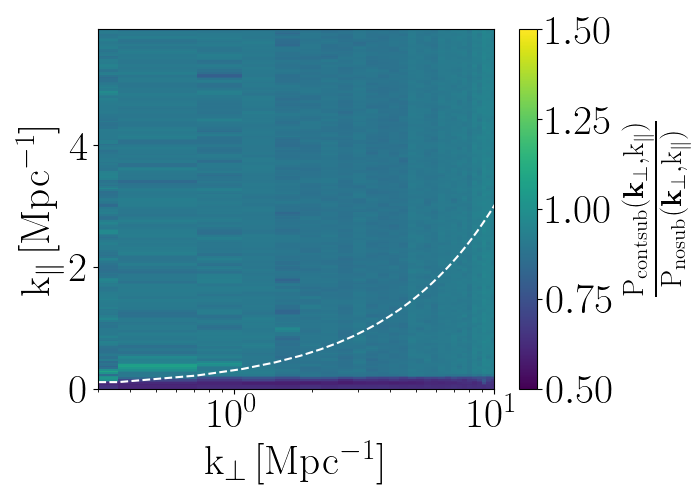}
\hspace{1pt}
\includegraphics[width=\columnwidth, height=2.3in]{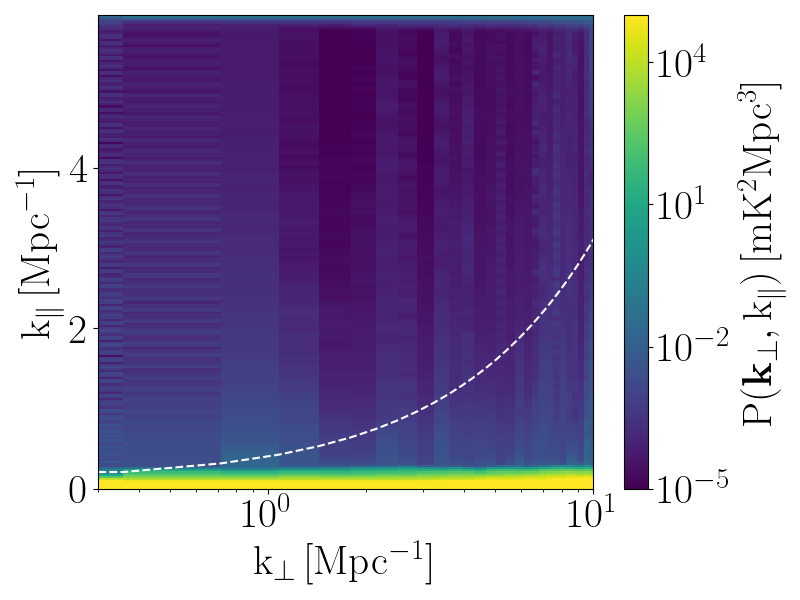}
\caption{\textbf{\textit{(left)}} Ratio of cylindrical auto-power spectra of continuum subtracted visibilities and visibilities pre-subtraction, averaged for all pointings. The ratio in the foreground wedge is $<1$, indicating a significant removal of continuum power. The region beyond the white dashed line shows a value of $\sim$1, indicating the measurement window is unaffected by continuum subtraction. \textbf{\textit{(right)}} Cylindrical power of the continuum model. The power in the measurement window (beyond the white dashed region) is $< 10^{-2}$\,mK$^{2}$Mpc$^{3}$, indicating negligible power present in the model outside the wedge.}
\label{signal_loss_cont}
\end{figure*}

The left panel of Figure \ref{signal_loss_cont} shows the ratio between the cylindrical power spectra of the continuum-subtracted visibilities and that of visibilities pre-subtraction. Any significant signal removal due to subtraction would result in the ratio being $<<$1 in the \hi measurement window. However, the ratio is almost consistently 1 in the region beyond the white dashed line, with the smallest value in the foreground wedge, where most of the removed power resides. The right panel of Figure \ref{signal_loss_cont} shows the power spectrum of the subtracted continuum model. We see here that inside the foreground wedge, the power spectrum has a value $\gtrsim 10^{5}$\,mK$^{2}$Mpc$^{3}$. Beyond the horizon limit (beyond the white dashed line), the values are very small ($< 10^{-2}$\,mK$^{2}$Mpc$^{3}$), a much smaller value than that expected for \hi signal.

Continuum subtraction removes a sky model dominated by the spectrally smooth foregrounds but preserves the spectral behaviour of \hi sources. This is essential for both intensity mapping and galaxy science. The method has already been tested for MIGHTEE data used for \hi studies (for example in \citealt{Sinigaglia_2022, ian_2024}). Figure \ref{signal_loss_cont} shows that the emission removed is almost exclusively confined to the wedge, which is evidence that the sky model used for the visibility domain continuum subtraction has well-modelled emission from the wedge only. It works well in reducing the foreground emission, without significantly impacting the underlying \hi.

From our tests, we could not find any evidence for \hi signal loss through the continuum removal process. Generally, the full extent of \hi signal loss can only be fully quantified through end-to-end simulations, which is beyond the scope of this paper but should be performed for future analysis. 

\section{Results}
\label{results}

In this section, we discuss the results obtained from the MIGHTEE-COSMOS data using the sub-band centred at 986\,MHz (z$\sim$0.44). We average all 15 pointings in the power spectrum domain and expect lower signal-to-noise than a single pointing with the same integration time. Nevertheless, owing to the excellent data quality and high-frequency resolution, this data is used to set the first upper limits on the HI power spectrum using the MIGHTEE survey. We choose the frequency band centred at z$\sim$0.44 and the scales are between $\sim$0.5\,Mpc$^{-1}$ to $\sim$10\,Mpc$^{-1}$ to avoid regions with very high noise.

We present the cross-correlation and auto-correlation power spectra obtained from the gridded visibilities. The latter acts as a conservative upper limit, as it is more likely to contain additive observational systematics. The cross-correlation amplitude is independent of the uncorrelated noise bias. However, negative systematics could artificially suppress the amplitude of the power spectrum, as outlined in \citep{morales_2023}. Comparing auto and cross-correlations, we verify the impact of noise and systematics and report consistent upper limits on the \hi power spectrum within the limited SNR of the MIGHTEE data.

\begin{figure}
\centering
\includegraphics[width=\columnwidth, height=7cm, trim={0.1cm 0.1cm 0.1cm 0cm},clip]{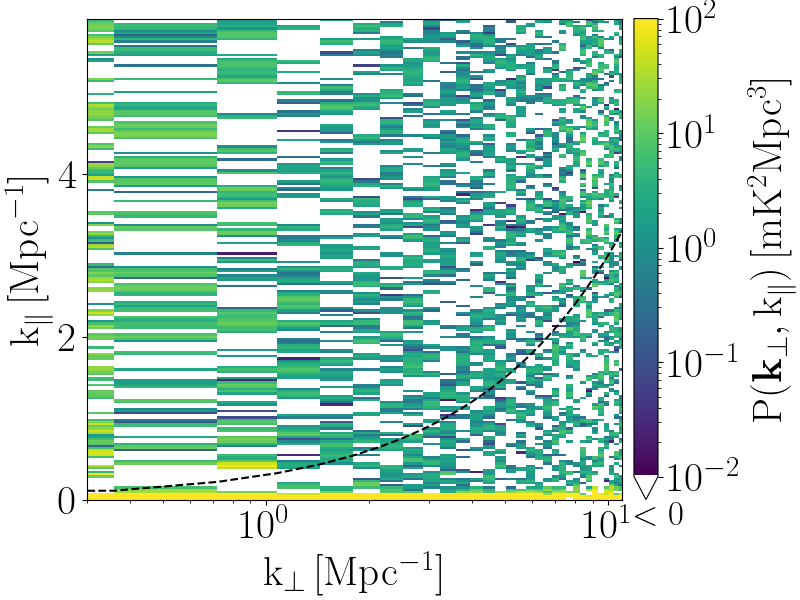}
\caption{Incoherently averaged cylindrical cross-power spectrum using continuum subtracted data for all pointings. The cross-power per pointing is the cross-correlation of alternate target scans to remove noise bias, which are then averaged. The black dashed curve shows the horizon limit of ${k}_{\parallel} \sim 0.3 {k}_{\perp}$, which gives the measurement window for the 1D power.}
\label{2dps}
\end{figure}

The power is calculated separately for each pointing and incoherently averaged using Equation \ref{ic_eq}. Figure \ref{2dps} shows the cylindrical cross-power spectrum for the combined MIGHTEE-COSMOS data centred at z$\sim$0.44. Owing to the fine frequency resolution of MIGHTEE, we can probe higher ${k}_{\parallel}$ modes compared to DEEP2 \citep{paul2023detection}. It is seen that the measurement window beyond the horizon limit (black dashed curve) does not show any dominant systematic contribution. The cross-correlation also removes some of the systematic contributions seen at the lower ${k}_{\perp}$ modes in Figure \ref{2dps_auto}. The foreground wedge region is less pronounced due to continuum subtraction.

\begin{figure}
\centering
\includegraphics[width=\columnwidth, height=7cm]{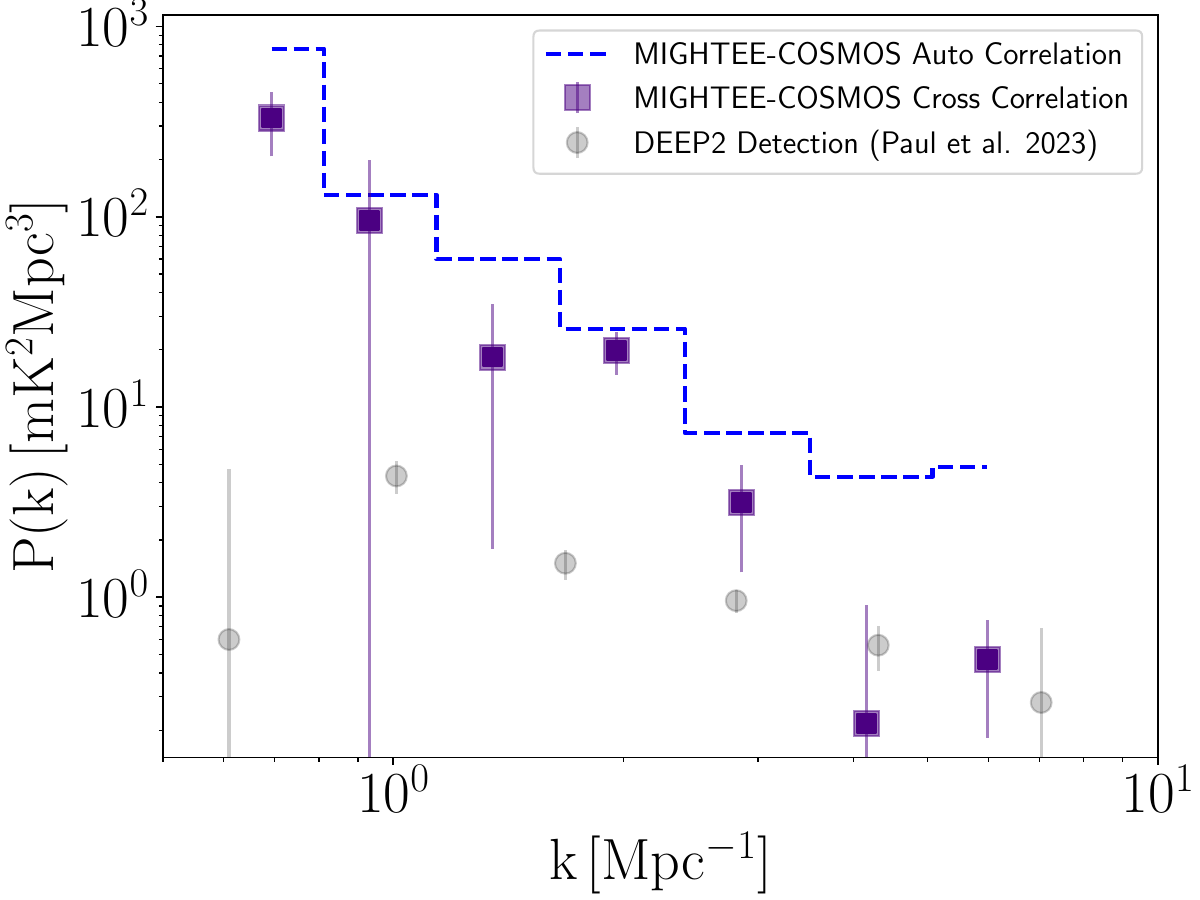}
\caption{Incoherently averaged 1D power spectrum from the MIGHTEE-COSMOS data at z$\sim$0.44. The measurements use modes beyond the horizon limit, i.e. the black dashed line in Figure \ref{2dps}. The indigo squares show the cross-power spectrum, and the blue dashed curve shows the auto-power spectrum. The auto-power spectrum is set as the absolute upper limit on the \hi power spectrum obtained in this work. The values for the \hi power obtained in \citep{paul2023detection} at z$\sim$0.44 are shown in grey circles for comparison.}
\label{1dps}
\end{figure}

We use the foreground avoidance method for calculating the 1D power spectrum averaging data in the region above the horizon limit ${k}_{\parallel} \sim 0.3 {k}_{\perp}$ (black dashed line in Figure \ref{2dps}). We compute the 3D power spectra in 7 logarithmic $k$ bins. Owing to the the smallest $k_{\perp}$ bin being highly noisy, we also exclude ${k}_{\perp}\leq$0.5\,Mpc$^{-1}$. The final incoherently averaged 1D power spectra at z$\sim$0.44 for the MIGHTEE-COSMOS observation is shown in Figure \ref{1dps}. The indigo squares indicate the cross-power spectrum, and blue dashed line marks the amplitude of the noise-bias removed auto-power spectrum. We also show the spherical power spectrum amplitude at z$\sim$0.44 reported in \citet{paul2023detection} by grey circles. The measured values of the power spectra are also shown in Table \ref{ps_values}, between $0.5\,\textrm{Mpc}^{-1} \lesssim k \lesssim 10\,\textrm{Mpc}^{-1}$.

We find that the values of the auto and cross-power are consistent within 2$\sigma$, except for the last two $k$ bins, which are generally noise dominated, as can be seen in the right panel of Figure \ref{2dps_auto}. We think the discrepancy is caused by noise and residual systematics in the auto-power spectrum in these bins, rather than a suppression in the amplitude of the cross-power. The consistency of the results demonstrates the robustness of our measurements, since the auto and cross-correlations are sensitive to different systematic and noise effects.

The measured cross-power spectrum shows a drop with increasing $k$, which is expected from the combination of shot noise and attenuation by the Finger-of-God (FoG) at high ${k}_{\parallel}$. Compared to the detection obtained in \citet{paul2023detection}, the power measured at a similar range of $k$ values is higher in this work. This is due to the lower SNR of the data as our analysis combines 15 overlapping pointings via incoherently averaging, whereas \citet{paul2023detection} analysed one deep pointing.

\begin{table}
\caption{\hi power spectrum constraints at z$\sim$0.44 obtained from MIGHTEE-COSMOS data after incoherent averaging all pointings. The measurement and errors are tabulated for the cross-power spectrum, while the upper limits indicate the values obtained from the bias-subtracted auto-power spectrum. All values are for the same $k$ bins.}
    \centering    
    \begin{tabular}{|c|c|c|c|c|}
    \hline
    \hline
    k & P(k) & $\sigma_{P}$ & P(k)/$\sigma_{P}$ & Upper Limit \\
    (Mpc$^{-1}$) & (mk$^2$Mpc$^3$) & & (mk$^2$Mpc$^3$) & (mk$^2$Mpc$^3$)  \\
    \hline
    0.69 & 330.14 & 122.46 & 2.69 & 764.50  \\
    0.93 & 95.72 & 103.44 & 0.93 & 130.21  \\
    1.35 & 18.29 & 16.49 & 1.11 & 59.67 \\
    1.96 & 19.79 & 5.00 & 3.96 & 25.82 \\
    2.85 & 3.15 & 1.80 & 1.75 & 7.29 \\
    4.16 & 0.22 & 0.69 & 0.32 & 4.30 \\
    5.99 & 0.47 & 0.29 & 1.63 & 4.83 \\
  \hline
     \hline
    \end{tabular}
    \label{ps_values}
\end{table}

\section{Summary}
\label{summary}
In this work, we demonstrate the capability for measuring the \hi IM signal on quasi-linear scales from the visibilities of DR-1 MIGHTEE spectral line observations \citep{ian_2024}. We use a small sub-band centred at redshift z$\sim$0.44 covering around 4 square degrees of the COSMOS field. COSMOS is a well-known extragalactic deep field with a rich population of interesting radio sources for continuum science. However, in our context, these sources contribute to the foreground emission which could potentially contaminate the \hi signal. In this work, we found that these emissions are well confined within the horizon scales of the wedge due to the high calibration accuracies, thus validating our approach of foreground avoidance in the power spectrum estimation. We compute both the auto-correlation and the cross-correlation powers from interleaving scans for individual pointings to identify noise and systematic effects. 
Using cross correlation, we derive the upper limit of 29.8 \,mK$^{2}$Mpc${^3}$ from the 2$\sigma$ value which is consistent with the auto-correlation value of 25.82\,mK$^{2}$Mpc${^3}$ on $k\sim$2\,Mpc$^{-1}$ on the \hi power spectrum. The results are within two orders of magnitude of \citet{paul2023detection} at the same redshift, even at the smallest $k$ values (despite significantly higher noise equivalent).

A few important things need to be considered in this context:

\begin{itemize}
    \item The data spans a large area over multiple pointings centred at slightly different right ascensions and declinations. Thus, the averaging is done in the power spectrum domain rather than in the visibility domain. Visibility averaging often makes residual errors easier to localise and remove, which cannot be done in this case. 
    
    \item The observations for MIGHTEE-COSMOS start late afternoon to late evening, none start late at night. Radio surveys are mostly done late at night to avoid ambient turbulence caused by the sun. However, inspection of the data showed that while some contamination is evident, the overall quality is not adversely affected.
   
    \item Direction-dependent effects have not been corrected for in this data set. We do not expect this to be an issue for the tightly packed mosaiced pointings used for the  L-band MIGHTEE-COSMOS observations. However, we will perform these corrections when combining data for other MIGHTEE fields with wider pointing spreads. 
    
     \item To assess the quality and usability of the data for IM, several tests were done as discussed in Section \ref{usability}. The main conclusion from these tests is that residual systematics can vary considerably between pointings (and even from one scan to the other in the same pointing). However, on combination and cross-correlation, most systematic contributions reduce considerably (Figure \ref{2dps}).  
     
    \item We select the $k$ modes conservatively, excluding those potentially containing systematics obscured by very high noise. 

    \item The limiting issue of identifying systematics due to lower signal-to-noise for this dataset will be addressed in future works incorporating more data.
    \end{itemize}

\section*{Acknowledgements}
The authors acknowledge Ian Heywood for the initial data calibration. The authors also thank the anonymous referees for their insightful comments that helped improve the quality of this paper. AM thanks the UK Research and Innovation Future Leaders Fellowship for supporting this research [grant MR/V026437/1]. LW is a UK Research and Innovation Future Leaders Fellow [grant MR/V026437/1]. MGS acknowledges support from the South African Radio Astronomy Observatory and National Research Foundation (Grant No. 84156). MJJ acknowledge generous support from the Hintze Family Charitable Foundation through the Oxford Hintze Centre for Astrophysical Surveys, support of the STFC consolidated grant [ST/S000488/1] and [ST/W000903/1] and from a UKRI Frontiers Research Grant [EP/X026639/1]. The MeerKAT telescope is operated by the South African Radio Astronomy Observatory, which is a facility of the National Research Foundation, an agency of the Department of Science and Innovation. We acknowledge the use of the Ilifu cloud computing facility (\url{https://www.ilifu.ac.za/}), a partnership between the University of Cape Town, the University of the Western Cape, Stellenbosch University, Sol Plaatje University and the Cape Peninsula University of Technology. The Ilifu facility is supported by contributions from the Inter-University Institute for Data Intensive Astronomy (IDIA – a partnership between the University of Cape Town, the University of Pretoria and the University of the Western Cape), the Computational Biology division at UCT and the Data Intensive Research Initiative of South Africa (DIRISA). The authors acknowledge the Centre for High Performance Computing (CHPC), South Africa, for providing computational resources to this research project. The authors thank the reviewer and the scientific editor for their insightful comments that have helped in improving this work. AM also thanks Sumanjit Chakraborty, Steve Cunnington, Keith Grainge and Amadeus Wild for helpful discussions. 

\section*{Data Availability}
The visibility data are available from the SARAO archive (\url{https://keycloak.sarao.ac.za/auth/realms/SKASA/login-actions/authenticate?client_id=archive.sarao.ac.za&tab_id=_BjAC2k2F90}) by searching for the capture block IDs listed in \autoref{table:pointings}. 
\\
\\
{\bf $Software$ :} \\
This work relies on the Python programming language (\url{https://www.python.org/}). The packages used here are \texttt{astropy} (\url{https://www.astropy.org/}; \citealt{astropy:2013,astropy:2018}), \texttt{numpy} (\url{https://numpy.org/}), scipy (\url{https://www.scipy.org/}), \texttt{matplotlib} (\url{https://matplotlib.org/}), \texttt{Common Astronomy Software Applications (CASA)}(\url{https://casaguides.nrao.edu/index.php?title=Main_Page}). Some analyses were done using \texttt{hiimtool} (\url{https://github.com/zhaotingchen/hiimtool}).



\bibliographystyle{mnras}
\bibliography{references} 

\begin{thebibliography}{}
\makeatletter
\relax
\def\mn@urlcharsother{\let\do\@makeother \do\$\do\&\do\#\do\^\do\_\do\%\do\~}
\def\mn@doi{\begingroup\mn@urlcharsother \@ifnextchar [ {\mn@doi@} {\mn@doi@[]}}
\def\mn@doi@[#1]#2{\def\@tempa{#1}\ifx\@tempa\@empty \href {http://dx.doi.org/#2} {doi:#2}\else \href {http://dx.doi.org/#2} {#1}\fi \endgroup}
\def\mn@eprint#1#2{\mn@eprint@#1:#2::\@nil}
\def\mn@eprint@arXiv#1{\href {http://arxiv.org/abs/#1} {{\tt arXiv:#1}}}
\def\mn@eprint@dblp#1{\href {http://dblp.uni-trier.de/rec/bibtex/#1.xml} {dblp:#1}}
\def\mn@eprint@#1:#2:#3:#4\@nil{\def\@tempa {#1}\def\@tempb {#2}\def\@tempc {#3}\ifx \@tempc \@empty \let \@tempc \@tempb \let \@tempb \@tempa \fi \ifx \@tempb \@empty \def\@tempb {arXiv}\fi \@ifundefined {mn@eprint@\@tempb}{\@tempb:\@tempc}{\expandafter \expandafter \csname mn@eprint@\@tempb\endcsname \expandafter{\@tempc}}}

\bibitem[\protect\citeauthoryear{Abbott et~al.,}{Abbott et~al.}{2022}]{des_weak_lens}
Abbott T. M.~C.,  et~al., 2022, \mn@doi [Phys. Rev. D] {10.1103/PhysRevD.105.023520}, 105, 023520

\bibitem[\protect\citeauthoryear{Abdurashidova et~al.,}{Abdurashidova et~al.}{2022}]{hera}
Abdurashidova Z.,  et~al., 2022, \mn@doi [The Astrophysical Journal] {10.3847/1538-4357/ac1c78}, 925, 221

\bibitem[\protect\citeauthoryear{Adams, Bowler, Jarvis, Häußler, McLure, Bunker, Dunlop  \& Verma}{Adams et~al.}{2020}]{10.1093/mnras/staa687}
Adams N.~J.,  Bowler R. A.~A.,  Jarvis M.~J.,  Häußler B.,  McLure R.~J.,  Bunker A.,  Dunlop J.~S.,   Verma A.,  2020, \mn@doi [Monthly Notices of the Royal Astronomical Society] {10.1093/mnras/staa687}, 494, 1771

\bibitem[\protect\citeauthoryear{Adams, Bowler, Jarvis, Häußler  \& Lagos}{Adams et~al.}{2021}]{10.1093/mnras/stab1956}
Adams N.~J.,  Bowler R. A.~A.,  Jarvis M.~J.,  Häußler B.,   Lagos C. D.~P.,  2021, \mn@doi [Monthly Notices of the Royal Astronomical Society] {10.1093/mnras/stab1956}, 506, 4933

\bibitem[\protect\citeauthoryear{Alam et~al.,}{Alam et~al.}{2021}]{clustring}
Alam S.,  et~al., 2021, \mn@doi [Phys. Rev. D] {10.1103/PhysRevD.103.083533}, 103, 083533

\bibitem[\protect\citeauthoryear{Amiri et~al.,}{Amiri et~al.}{2023}]{chime_galaxy}
Amiri M.,  et~al., 2023, \mn@doi [The Astrophysical Journal] {10.3847/1538-4357/acb13f}, 947, 16

\bibitem[\protect\citeauthoryear{Anderson et~al.,}{Anderson et~al.}{2018}]{parkes_im}
Anderson C.~J.,  et~al., 2018, \mn@doi [Monthly Notices of the Royal Astronomical Society] {10.1093/mnras/sty346}, 476, 3382

\bibitem[\protect\citeauthoryear{{Astropy Collaboration} et~al.,}{{Astropy Collaboration} et~al.}{2013}]{astropy:2013}
{Astropy Collaboration} et~al., 2013, \mn@doi [\aap] {10.1051/0004-6361/201322068}, \href {http://adsabs.harvard.edu/abs/2013A%26A...558A..33A} {558, A33}

\bibitem[\protect\citeauthoryear{{Bandura} et~al.,}{{Bandura} et~al.}{2014}]{chime}
{Bandura} K.,  et~al., 2014, in {Stepp} L.~M.,  {Gilmozzi} R.,   {Hall} H.~J.,  eds,  Society of Photo-Optical Instrumentation Engineers (SPIE) Conference Series Vol. 9145, Ground-based and Airborne Telescopes V. p. 914522 (\mn@eprint {arXiv} {1406.2288}), \mn@doi{10.1117/12.2054950}

\bibitem[\protect\citeauthoryear{Barry, Hazelton, Sullivan, Morales  \& Pober}{Barry et~al.}{2016}]{barry16}
Barry N.,  Hazelton B.,  Sullivan I.,  Morales M.~F.,   Pober J.~C.,  2016, \mn@doi [Monthly Notices of the Royal Astronomical Society] {10.1093/mnras/stw1380}, 461, 3135

\bibitem[\protect\citeauthoryear{Battye, Davies  \& Weller}{Battye et~al.}{2004}]{battye_2004}
Battye R.~A.,  Davies R.~D.,   Weller J.,  2004, \mn@doi [Monthly Notices of the Royal Astronomical Society] {10.1111/j.1365-2966.2004.08416.x}, 355, 1339

\bibitem[\protect\citeauthoryear{{Bharadwaj}, {Nath}  \& {Sethi}}{{Bharadwaj} et~al.}{2001}]{sb_2001}
{Bharadwaj} S.,  {Nath} B.~B.,   {Sethi} S.~K.,  2001, \mn@doi [Journal of Astrophysics and Astronomy] {10.1007/BF02933588}, \href {https://ui.adsabs.harvard.edu/abs/2001JApA...22...21B} {22, 21}

\bibitem[\protect\citeauthoryear{Bowler, Jarvis, Dunlop, McLure, McLeod, Adams, Milvang-Jensen  \& McCracken}{Bowler et~al.}{2020}]{rebecca_2020}
Bowler R. A.~A.,  Jarvis M.~J.,  Dunlop J.~S.,  McLure R.~J.,  McLeod D.~J.,  Adams N.~J.,  Milvang-Jensen B.,   McCracken H.~J.,  2020, \mn@doi [Monthly Notices of the Royal Astronomical Society] {10.1093/mnras/staa313}, 493, 2059

\bibitem[\protect\citeauthoryear{{CHIME Collaboration} et~al.,}{{CHIME Collaboration} et~al.}{2023}]{chime_lya}
{CHIME Collaboration} et~al., 2023, \mn@doi [arXiv e-prints] {10.48550/arXiv.2309.04404}, \href {https://ui.adsabs.harvard.edu/abs/2023arXiv230904404C} {p. arXiv:2309.04404}

\bibitem[\protect\citeauthoryear{Chakraborty et~al.,}{Chakraborty et~al.}{2021}]{Chakraborty_2021}
Chakraborty A.,  et~al., 2021, \mn@doi [The Astrophysical Journal Letters] {10.3847/2041-8213/abd17a}, 907, L7

\bibitem[\protect\citeauthoryear{Chan et~al.,}{Chan et~al.}{2022}]{des_clustering}
Chan K.~C.,  et~al., 2022, \mn@doi [Phys. Rev. D] {10.1103/PhysRevD.106.123502}, 106, 123502

\bibitem[\protect\citeauthoryear{Chang, Pen, Peterson  \& McDonald}{Chang et~al.}{2008}]{PhysRevLett.100.091303}
Chang T.-C.,  Pen U.-L.,  Peterson J.~B.,   McDonald P.,  2008, \mn@doi [Phys. Rev. Lett.] {10.1103/PhysRevLett.100.091303}, 100, 091303

\bibitem[\protect\citeauthoryear{{Chang}, {Pen}, {Bandura}  \& {Peterson}}{{Chang} et~al.}{2010}]{2010Natur.466..463C}
{Chang} T.-C.,  {Pen} U.-L.,  {Bandura} K.,   {Peterson} J.~B.,  2010, \mn@doi [\nat] {10.1038/nature09187}, \href {https://ui.adsabs.harvard.edu/abs/2010Natur.466..463C} {466, 463}

\bibitem[\protect\citeauthoryear{Chapman et~al.,}{Chapman et~al.}{2015}]{Chapman2015}
Chapman E.,  et~al., 2015, in Advancing Astrophysics with the Square Kilometre Array (AASKA14). p.~5, \url {https://ui.adsabs.harvard.edu/abs/2015aska.confE...5C}

\bibitem[\protect\citeauthoryear{Chen, Wolz, Spinelli  \& Murray}{Chen et~al.}{2021}]{zhaoting_theo}
Chen Z.,  Wolz L.,  Spinelli M.,   Murray S.~G.,  2021, \mn@doi [Monthly Notices of the Royal Astronomical Society] {10.1093/mnras/stab386}, 502, 5259

\bibitem[\protect\citeauthoryear{Chen, Wolz  \& Battye}{Chen et~al.}{2022}]{zchen_1}
Chen Z.,  Wolz L.,   Battye R.,  2022, \mn@doi [Monthly Notices of the Royal Astronomical Society] {10.1093/mnras/stac3288}, 518, 2971

\bibitem[\protect\citeauthoryear{Chen, Chapman, Wolz  \& Mazumder}{Chen et~al.}{2023}]{zchen_ska}
Chen Z.,  Chapman E.,  Wolz L.,   Mazumder A.,  2023, \mn@doi [Monthly Notices of the Royal Astronomical Society] {10.1093/mnras/stad2102}, 524, 3724

\bibitem[\protect\citeauthoryear{Cleary et~al.,}{Cleary et~al.}{2022}]{CO}
Cleary K.~A.,  et~al., 2022, \mn@doi [The Astrophysical Journal] {10.3847/1538-4357/ac63cc}, 933, 182

\bibitem[\protect\citeauthoryear{Cunnington, Wolz, Pourtsidou  \& Bacon}{Cunnington et~al.}{2019}]{steve_2019}
Cunnington S.,  Wolz L.,  Pourtsidou A.,   Bacon D.,  2019, \mn@doi [Monthly Notices of the Royal Astronomical Society] {10.1093/mnras/stz1916}, 488, 5452

\bibitem[\protect\citeauthoryear{Cunnington, Irfan, Carucci, Pourtsidou  \& Bobin}{Cunnington et~al.}{2021}]{steve_2021}
Cunnington S.,  Irfan M.~O.,  Carucci I.~P.,  Pourtsidou A.,   Bobin J.,  2021, \mn@doi [Monthly Notices of the Royal Astronomical Society] {10.1093/mnras/stab856}, 504, 208

\bibitem[\protect\citeauthoryear{Cunnington et~al.,}{Cunnington et~al.}{2022}]{steve_2022}
Cunnington S.,  et~al., 2022, \mn@doi [Monthly Notices of the Royal Astronomical Society] {10.1093/mnras/stac3060}, 518, 6262

\bibitem[\protect\citeauthoryear{{DESI Collaboration} et~al.,}{{DESI Collaboration} et~al.}{2022}]{desi}
{DESI Collaboration} et~al., 2022, \mn@doi [The Astronomical Journal] {10.3847/1538-3881/ac882b}, 164, 207

\bibitem[\protect\citeauthoryear{Datta, Bowman  \& Carilli}{Datta et~al.}{2010}]{Datta2010}
Datta A.,  Bowman J.~D.,   Carilli C.~L.,  2010, \mn@doi [The Astrophysical Journal] {10.1088/0004-637x/724/1/526}, 724, 526

\bibitem[\protect\citeauthoryear{Dawson et~al.,}{Dawson et~al.}{2016}]{Dawson_2016}
Dawson K.~S.,  et~al., 2016, \mn@doi [The Astronomical Journal] {10.3847/0004-6256/151/2/44}, 151, 44

\bibitem[\protect\citeauthoryear{Drinkwater et~al.,}{Drinkwater et~al.}{2010}]{wigglez}
Drinkwater M.~J.,  et~al., 2010, \mn@doi [Monthly Notices of the Royal Astronomical Society] {10.1111/j.1365-2966.2009.15754.x}, 401, 1429

\bibitem[\protect\citeauthoryear{Driver et~al.,}{Driver et~al.}{2011}]{gama}
Driver S.~P.,  et~al., 2011, \mn@doi [Monthly Notices of the Royal Astronomical Society] {10.1111/j.1365-2966.2010.18188.x}, 413, 971

\bibitem[\protect\citeauthoryear{Ewall-Wice, Dillon, Liu  \& Hewitt}{Ewall-Wice et~al.}{2017}]{ewall2017}
Ewall-Wice A.,  Dillon J.~S.,  Liu A.,   Hewitt J.,  2017, \mn@doi [Monthly Notices of the Royal Astronomical Society] {10.1093/mnras/stx1221}, 470, 1849

\bibitem[\protect\citeauthoryear{Furlanetto, {Peng Oh}  \& Briggs}{Furlanetto et~al.}{2006}]{Furlanetto2006}
Furlanetto S.~R.,  {Peng Oh} S.,   Briggs F.~H.,  2006, \mn@doi [Physics Reports] {https://doi.org/10.1016/j.physrep.2006.08.002}, 433, 181

\bibitem[\protect\citeauthoryear{Goedhart}{Goedhart}{2020a}]{Specs}
Goedhart S.,  2020a, MeerKAT specifications, \url {https://skaafrica.atlassian.net/wiki/spaces/ESDKB/pages/277315585/MeerKAT+specifications}

\bibitem[\protect\citeauthoryear{Goedhart}{Goedhart}{2020b}]{flags}
Goedhart S.,  2020b, Radio Frequency Interference (RFI), \url {https://skaafrica.atlassian.net/wiki/spaces/ESDKB/pages/305332225/Radio+Frequency+Interference+RFI}

\bibitem[\protect\citeauthoryear{Hale et~al.,}{Hale et~al.}{2024}]{hale_2024}
Hale C.~L.,  et~al., 2024, \mn@doi [Monthly Notices of the Royal Astronomical Society] {10.1093/mnras/stae2528}, p. stae2528

\bibitem[\protect\citeauthoryear{Heywood et~al.,}{Heywood et~al.}{2021}]{ian_2021}
Heywood I.,  et~al., 2021, \mn@doi [Monthly Notices of the Royal Astronomical Society] {10.1093/mnras/stab3021}, 509, 2150

\bibitem[\protect\citeauthoryear{Heywood et~al.,}{Heywood et~al.}{2024}]{ian_2024}
Heywood I.,  et~al., 2024, \mn@doi [Monthly Notices of the Royal Astronomical Society] {10.1093/mnras/stae2081}, 534, 76

\bibitem[\protect\citeauthoryear{{Hugo}, {Perkins}, {Merry}, {Mauch}  \& {Smirnov}}{{Hugo} et~al.}{2022}]{tricolor}
{Hugo} B.~V.,  {Perkins} S.,  {Merry} B.,  {Mauch} T.,   {Smirnov} O.~M.,  2022, in {Ruiz} J.~E.,  {Pierfedereci} F.,   {Teuben} P.,  eds,  Astronomical Society of the Pacific Conference Series Vol. 532, Astronomical Society of the Pacific Conference Series. p.~541 (\mn@eprint {arXiv} {2206.09179}), \mn@doi{10.48550/arXiv.2206.09179}

\bibitem[\protect\citeauthoryear{Jarvis et~al.,}{Jarvis et~al.}{2016}]{mightee}
Jarvis M.,  et~al., 2016, in Proceedings of MeerKAT Science: On the Pathway to the SKA {\textemdash} PoS(MeerKAT2016). p.~006, \mn@doi{10.22323/1.277.0006}

\bibitem[\protect\citeauthoryear{Jeli{\'c} et~al.,}{Jeli{\'c} et~al.}{2008}]{jelic2008}
Jeli{\'c} V.,  et~al., 2008, \mn@doi [MNRAS] {10.1111/j.1365-2966.2008.13634.x}, 389, 1319

\bibitem[\protect\citeauthoryear{Jonas}{Jonas}{2018}]{Jonas:2018Jr}
Jonas J.,  2018, in Proceedings of MeerKAT Science: On the Pathway to the SKA {\textemdash} PoS(MeerKAT2016). p.~001, \mn@doi{10.22323/1.277.0001}

\bibitem[\protect\citeauthoryear{{Kovetz} et~al.,}{{Kovetz} et~al.}{2017}]{kovetz2017lineintensity}
{Kovetz} E.~D.,  et~al., 2017, \mn@doi [arXiv e-prints] {10.48550/arXiv.1709.09066}, \href {https://ui.adsabs.harvard.edu/abs/2017arXiv170909066K} {p. arXiv:1709.09066}

\bibitem[\protect\citeauthoryear{Laporte et~al.,}{Laporte et~al.}{2017}]{OIII}
Laporte N.,  et~al., 2017, \mn@doi [The Astrophysical Journal Letters] {10.3847/2041-8213/aa62aa}, 837, L21

\bibitem[\protect\citeauthoryear{Liu, Parsons  \& Trott}{Liu et~al.}{2014}]{liu2014}
Liu A.,  Parsons A.~R.,   Trott C.~M.,  2014, \mn@doi [Phys. Rev. D] {10.1103/PhysRevD.90.023018}, 90, 023018

\bibitem[\protect\citeauthoryear{Liu, Zhang  \& Parsons}{Liu et~al.}{2016}]{Liu_2016}
Liu A.,  Zhang Y.,   Parsons A.~R.,  2016, \mn@doi [The Astrophysical Journal] {10.3847/1538-4357/833/2/242}, 833, 242

\bibitem[\protect\citeauthoryear{Masui et~al.,}{Masui et~al.}{2013}]{Masui_2013}
Masui K.~W.,  et~al., 2013, \mn@doi [The Astrophysical Journal Letters] {10.1088/2041-8205/763/1/L20}, 763, L20

\bibitem[\protect\citeauthoryear{Matshawule, Spinelli, Santos  \& Ngobese}{Matshawule et~al.}{2021}]{pb_im}
Matshawule S.~D.,  Spinelli M.,  Santos M.~G.,   Ngobese S.,  2021, \mn@doi [Monthly Notices of the Royal Astronomical Society] {10.1093/mnras/stab1688}, 506, 5075

\bibitem[\protect\citeauthoryear{Mazumder, Datta, Chakraborty  \& Majumdar}{Mazumder et~al.}{2022}]{aishrila2022}
Mazumder A.,  Datta A.,  Chakraborty A.,   Majumdar S.,  2022, \mn@doi [Monthly Notices of the Royal Astronomical Society] {10.1093/mnras/stac1994}, 515, 4020

\bibitem[\protect\citeauthoryear{Mertens et~al.,}{Mertens et~al.}{2020}]{lofar2}
Mertens F.~G.,  et~al., 2020, \mn@doi [Monthly Notices of the Royal Astronomical Society] {10.1093/mnras/staa327}, 493, 1662

\bibitem[\protect\citeauthoryear{Morales \& Hewitt}{Morales \& Hewitt}{2004}]{Morales_hewitt}
Morales M.~F.,  Hewitt J.,  2004, \mn@doi [The Astrophysical Journal] {10.1086/424437}, 615, 7

\bibitem[\protect\citeauthoryear{Morales \& Wyithe}{Morales \& Wyithe}{2010}]{morales2010}
Morales M.~F.,  Wyithe J. S.~B.,  2010, \mn@doi [Annual Review of Astronomy and Astrophysics] {10.1146/annurev-astro-081309-130936}, 48, 127

\bibitem[\protect\citeauthoryear{Morales, Hazelton, Sullivan  \& Beardsley}{Morales et~al.}{2012}]{Morales2012}
Morales M.~F.,  Hazelton B.,  Sullivan I.,   Beardsley A.,  2012, \mn@doi [The Astrophysical Journal] {10.1088/0004-637x/752/2/137}, 752, 137

\bibitem[\protect\citeauthoryear{Morales, Beardsley, Pober, Barry, Hazelton, Jacobs  \& Sullivan}{Morales et~al.}{2018}]{morales_2019}
Morales M.~F.,  Beardsley A.,  Pober J.,  Barry N.,  Hazelton B.,  Jacobs D.,   Sullivan I.,  2018, \mn@doi [Monthly Notices of the Royal Astronomical Society] {10.1093/mnras/sty2844}, 483, 2207

\bibitem[\protect\citeauthoryear{Morales, Pober  \& Hazelton}{Morales et~al.}{2023}]{morales_2023}
Morales M.~F.,  Pober J.,   Hazelton B.~J.,  2023, \mn@doi [Monthly Notices of the Royal Astronomical Society] {10.1093/mnras/stad2357}, 525, 2834

\bibitem[\protect\citeauthoryear{Offringa, de Bruyn, Biehl, Zaroubi, Bernardi  \& Pandey}{Offringa et~al.}{2010}]{offringa2010}
Offringa A.~R.,  de Bruyn A.~G.,  Biehl M.,  Zaroubi S.,  Bernardi G.,   Pandey V.~N.,  2010, \mn@doi [Monthly Notices of the Royal Astronomical Society] {10.1111/j.1365-2966.2010.16471.x}, 405, 155

\bibitem[\protect\citeauthoryear{Paciga et~al.,}{Paciga et~al.}{2011}]{pacgia}
Paciga G.,  et~al., 2011, Monthly Notices of the Royal Astronomical Society, 413, 1174

\bibitem[\protect\citeauthoryear{Parsons, Pober, McQuinn, Jacobs  \& Aguirre}{Parsons et~al.}{2012}]{Parsons2012}
Parsons A.,  Pober J.,  McQuinn M.,  Jacobs D.,   Aguirre J.,  2012, \mn@doi [The Astrophysical Journal] {10.1088/0004-637x/753/1/81}, 753, 81

\bibitem[\protect\citeauthoryear{Parsons et~al.,}{Parsons et~al.}{2014}]{Parsons_2014}
Parsons A.~R.,  et~al., 2014, \mn@doi [The Astrophysical Journal] {10.1088/0004-637X/788/2/106}, 788, 106

\bibitem[\protect\citeauthoryear{Paul, Santos, Townsend, Jarvis, Maddox, Collier, Frank  \& Taylor}{Paul et~al.}{2021}]{sourabh_2021}
Paul S.,  Santos M.~G.,  Townsend J.,  Jarvis M.~J.,  Maddox N.,  Collier J.~D.,  Frank B.~S.,   Taylor R.,  2021, \mn@doi [Monthly Notices of the Royal Astronomical Society] {10.1093/mnras/stab1089}, 505, 2039

\bibitem[\protect\citeauthoryear{{Paul}, {Santos}, {Chen}  \& {Wolz}}{{Paul} et~al.}{2023}]{paul2023detection}
{Paul} S.,  {Santos} M.~G.,  {Chen} Z.,   {Wolz} L.,  2023, \mn@doi [arXiv e-prints] {10.48550/arXiv.2301.11943}, \href {https://ui.adsabs.harvard.edu/abs/2023arXiv230111943P} {p. arXiv:2301.11943}

\bibitem[\protect\citeauthoryear{{Planck Collaboration} et~al.,}{{Planck Collaboration} et~al.}{2020}]{planck_cosmology}
{Planck Collaboration} et~al., 2020, \mn@doi [A&A] {10.1051/0004-6361/201833910}, 641, A6

\bibitem[\protect\citeauthoryear{Pourtsidou, Bacon  \& Crittenden}{Pourtsidou et~al.}{2017}]{10.1093/mnras/stx1479}
Pourtsidou A.,  Bacon D.,   Crittenden R.,  2017, \mn@doi [Monthly Notices of the Royal Astronomical Society] {10.1093/mnras/stx1479}, 470, 4251

\bibitem[\protect\citeauthoryear{{Price-Whelan} et~al.,}{{Price-Whelan} et~al.}{2018}]{astropy:2018}
{Price-Whelan} A.~M.,  et~al., 2018, \mn@doi [\aj] {10.3847/1538-3881/aabc4f}, \href {https://ui.adsabs.harvard.edu/#abs/2018AJ....156..123T} {156, 123}

\bibitem[\protect\citeauthoryear{Ratcliffe}{Ratcliffe}{2020}]{sdp}
Ratcliffe S.,  2020, SDP pipelines overview, \url {https://skaafrica.atlassian.net/wiki/spaces/ESDKB/pages/338723406/SDP+pipelines+overview}

\bibitem[\protect\citeauthoryear{SKA Cosmology~SWG Bacon et~al.,}{SKA Cosmology~SWG et~al.}{2020}]{red_book}
SKA Cosmology~SWG Bacon D.~J.,  et~al., 2020, \mn@doi [Publications of the Astronomical Society of Australia] {10.1017/pasa.2019.51}, 37, e007

\bibitem[\protect\citeauthoryear{{Santos} et~al.,}{{Santos} et~al.}{2016}]{santos2017meerklass}
{Santos} M.,  et~al., 2016, in MeerKAT Science: On the Pathway to the SKA. p.~32 (\mn@eprint {arXiv} {1709.06099}), \mn@doi{10.22323/1.277.0032}

\bibitem[\protect\citeauthoryear{Sinigaglia et~al.,}{Sinigaglia et~al.}{2022}]{Sinigaglia_2022}
Sinigaglia F.,  et~al., 2022, \mn@doi [The Astrophysical Journal Letters] {10.3847/2041-8213/ac85ae}, 935, L13

\bibitem[\protect\citeauthoryear{Spinelli, Zoldan, De Lucia, Xie  \& Viel}{Spinelli et~al.}{2020}]{marta_theo}
Spinelli M.,  Zoldan A.,  De Lucia G.,  Xie L.,   Viel M.,  2020, \mn@doi [Monthly Notices of the Royal Astronomical Society] {10.1093/mnras/staa604}, 493, 5434

\bibitem[\protect\citeauthoryear{Spinelli, Carucci, Cunnington, Harper, Irfan, Fonseca, Pourtsidou  \& Wolz}{Spinelli et~al.}{2021}]{sd_data_ch}
Spinelli M.,  Carucci I.~P.,  Cunnington S.,  Harper S.~E.,  Irfan M.~O.,  Fonseca J.,  Pourtsidou A.,   Wolz L.,  2021, \mn@doi [Monthly Notices of the Royal Astronomical Society] {10.1093/mnras/stab3064}, 509, 2048

\bibitem[\protect\citeauthoryear{Switzer et~al.,}{Switzer et~al.}{2013}]{gbt_ii}
Switzer E.~R.,  et~al., 2013, \mn@doi [Monthly Notices of the Royal Astronomical Society: Letters] {10.1093/mnrasl/slt074}, 434, L46

\bibitem[\protect\citeauthoryear{Switzer, Anderson, Pullen  \& Yang}{Switzer et~al.}{2019}]{Switzer_2019}
Switzer E.~R.,  Anderson C.~J.,  Pullen A.~R.,   Yang S.,  2019, \mn@doi [The Astrophysical Journal] {10.3847/1538-4357/aaf9ab}, 872, 82

\bibitem[\protect\citeauthoryear{{The Dark Energy Survey Collaboration}}{{The Dark Energy Survey Collaboration}}{2005}]{des}
{The Dark Energy Survey Collaboration} 2005, The Dark Energy Survey (\mn@eprint {arXiv} {astro-ph/0510346})

\bibitem[\protect\citeauthoryear{Trott \& Wayth}{Trott \& Wayth}{2016}]{trottwayth2016}
Trott C.~M.,  Wayth R.~B.,  2016, \mn@doi [Publications of the Astronomical Society of Australia] {10.1017/pasa.2016.18}, 33, e019

\bibitem[\protect\citeauthoryear{Trott, Wayth  \& Tingay}{Trott et~al.}{2012}]{Trott2012}
Trott C.~M.,  Wayth R.~B.,   Tingay S.~J.,  2012, \mn@doi [The Astrophysical Journal] {10.1088/0004-637x/757/1/101}, 757, 101

\bibitem[\protect\citeauthoryear{Trott et~al.,}{Trott et~al.}{2020}]{mwa2}
Trott C.~M.,  et~al., 2020, \mn@doi [Monthly Notices of the Royal Astronomical Society] {10.1093/mnras/staa414}, 493, 4711

\bibitem[\protect\citeauthoryear{Vedantham, Shankar  \& Subrahmanyan}{Vedantham et~al.}{2012}]{Vedantham2012}
Vedantham H.,  Shankar N.~U.,   Subrahmanyan R.,  2012, \mn@doi [The Astrophysical Journal] {10.1088/0004-637x/745/2/176}, 745, 176

\bibitem[\protect\citeauthoryear{Visbal, Trac  \& Loeb}{Visbal et~al.}{2011}]{Visbal_2011}
Visbal E.,  Trac H.,   Loeb A.,  2011, \mn@doi [Journal of Cosmology and Astroparticle Physics] {10.1088/1475-7516/2011/08/010}, 2011, 010

\bibitem[\protect\citeauthoryear{Wang et~al.,}{Wang et~al.}{2021}]{wang_meerklass}
Wang J.,  et~al., 2021, \mn@doi [Monthly Notices of the Royal Astronomical Society] {10.1093/mnras/stab1365}, 505, 3698

\bibitem[\protect\citeauthoryear{Whittam et~al.,}{Whittam et~al.}{2023}]{multi_w_cosmos}
Whittam I.~H.,  et~al., 2023, \mn@doi [Monthly Notices of the Royal Astronomical Society] {10.1093/mnras/stad3307}, 527, 3231

\bibitem[\protect\citeauthoryear{Wolz, Abdalla, Blake, Shaw, Chapman  \& Rawlings}{Wolz et~al.}{2014}]{wolz_2016}
Wolz L.,  Abdalla F.~B.,  Blake C.,  Shaw J.~R.,  Chapman E.,   Rawlings S.,  2014, \mn@doi [Monthly Notices of the Royal Astronomical Society] {10.1093/mnras/stu792}, 441, 3271

\bibitem[\protect\citeauthoryear{Wolz et~al.,}{Wolz et~al.}{2021}]{wolz_2021}
Wolz L.,  et~al., 2021, \mn@doi [Monthly Notices of the Royal Astronomical Society] {10.1093/mnras/stab3621}, 510, 3495

\bibitem[\protect\citeauthoryear{Wyithe, Loeb  \& Geil}{Wyithe et~al.}{2008}]{wyithe}
Wyithe J. S.~B.,  Loeb A.,   Geil P.~M.,  2008, \mn@doi [Monthly Notices of the Royal Astronomical Society] {10.1111/j.1365-2966.2007.12631.x}, 383, 1195

\bibitem[\protect\citeauthoryear{York et~al.,}{York et~al.}{2000}]{sdss}
York D.~G.,  et~al., 2000, \mn@doi [The Astronomical Journal] {10.1086/301513}, 120, 1579

\bibitem[\protect\citeauthoryear{Yue, Ferrara, Pallottini, Gallerani  \& Vallini}{Yue et~al.}{2015}]{CII}
Yue B.,  Ferrara A.,  Pallottini A.,  Gallerani S.,   Vallini L.,  2015, \mn@doi [Monthly Notices of the Royal Astronomical Society] {10.1093/mnras/stv933}, 450, 3829

\bibitem[\protect\citeauthoryear{{de Jong}, {Verdoes Kleijn}, {Kuijken}  \& {Valentijn}}{{de Jong} et~al.}{2013}]{Kids}
{de Jong} J. T.~A.,  {Verdoes Kleijn} G.~A.,  {Kuijken} K.~H.,   {Valentijn} E.~A.,  2013, \mn@doi [Experimental Astronomy] {10.1007/s10686-012-9306-1}, \href {https://ui.adsabs.harvard.edu/abs/2013ExA....35...25D} {35, 25}

\makeatother
\end{thebibliography}



\appendix
\section{Excess Power in Stokes V}
\label{stokes v}
\begin{figure*}
\centering
\includegraphics[width=0.75\textwidth, height=0.25\textwidth]{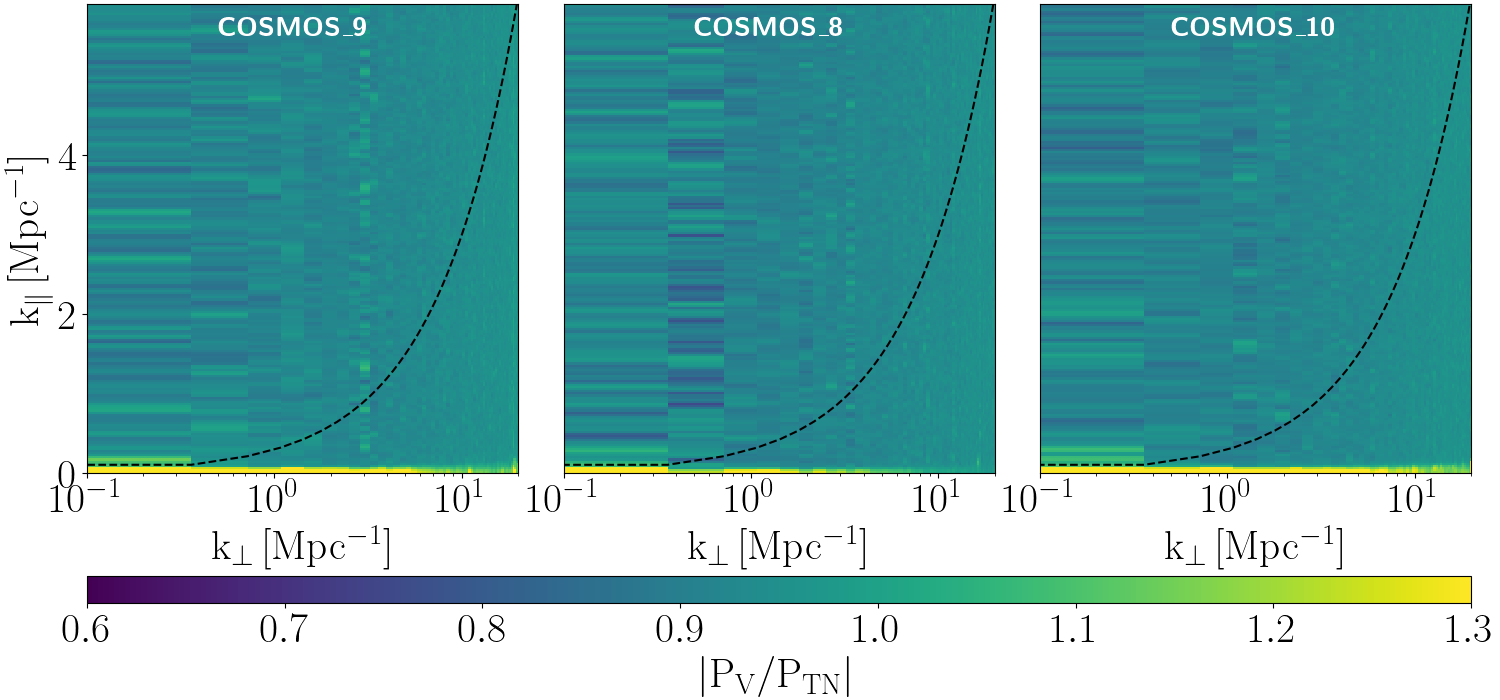}
\includegraphics[width=0.75\textwidth, height=0.25\textwidth]{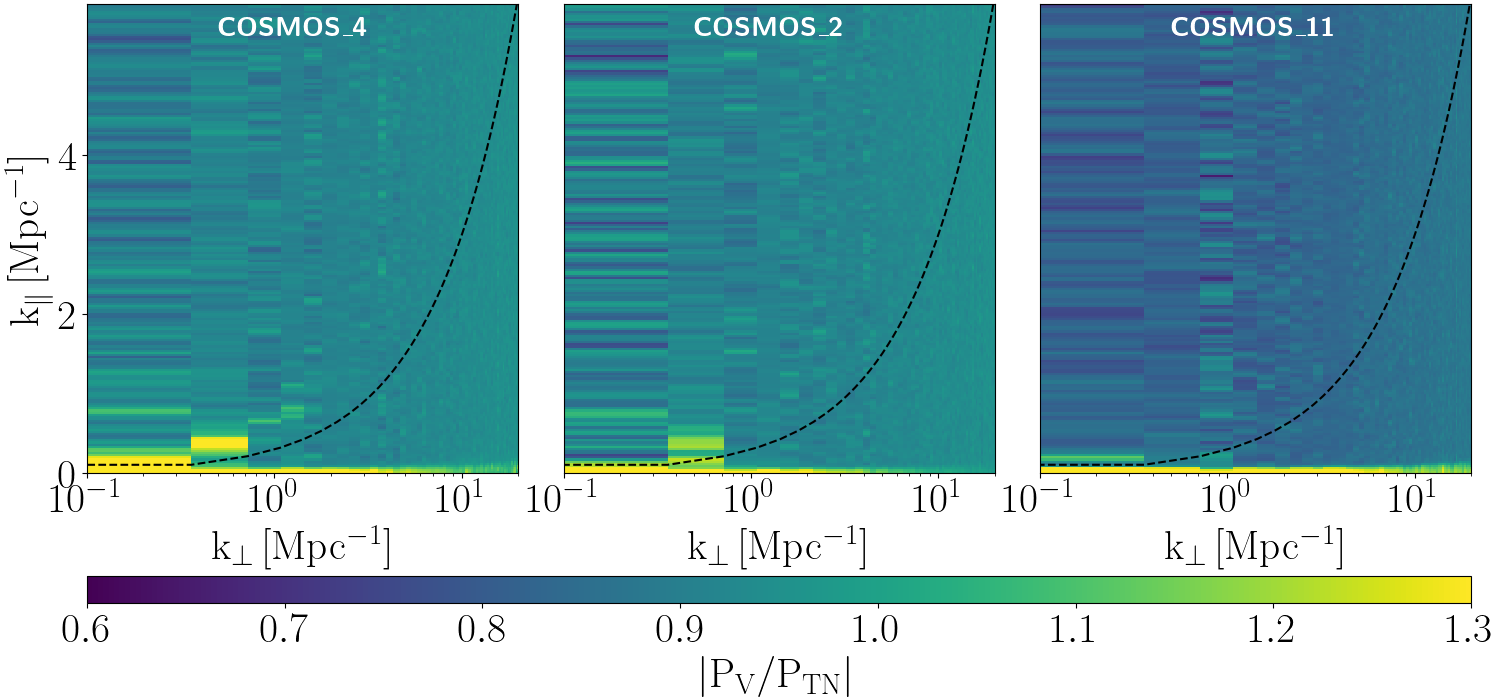}
\includegraphics[width=0.75\textwidth, height=0.25\textwidth]{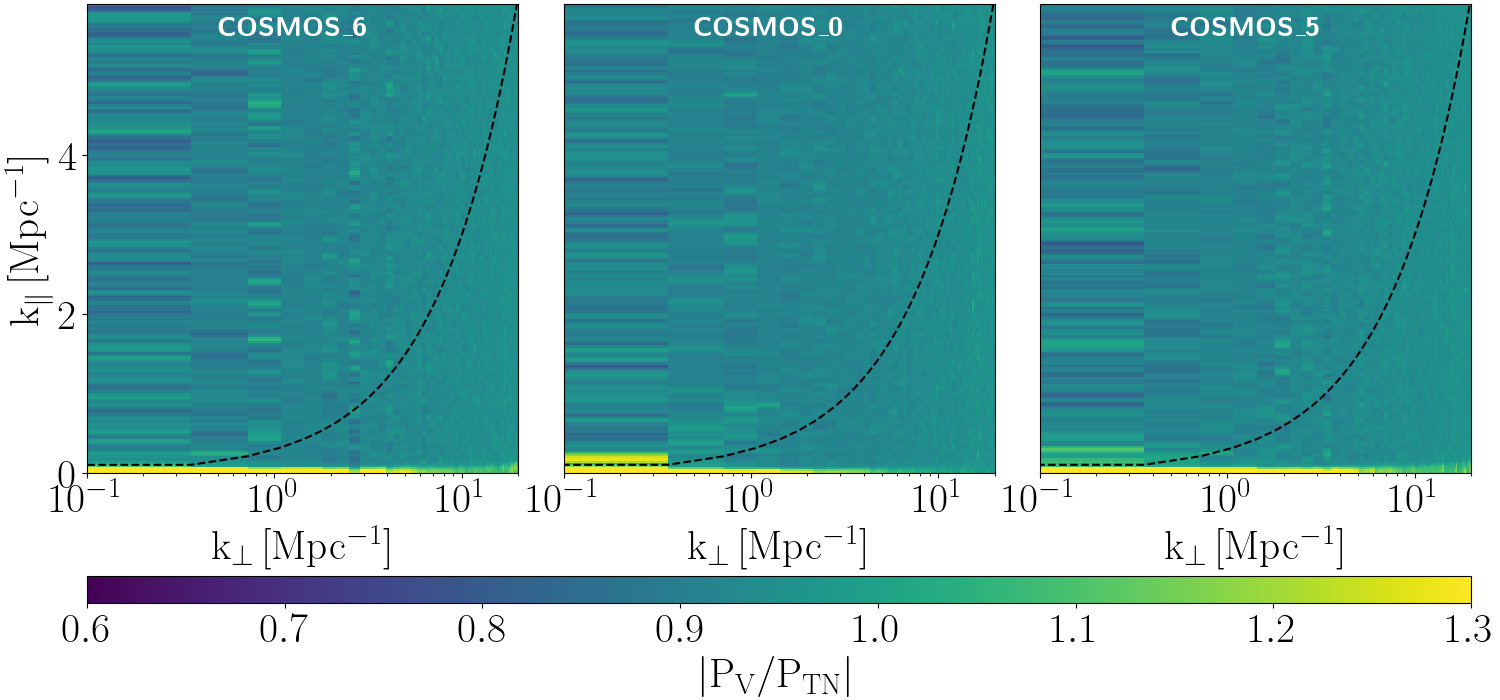}
\includegraphics[width=0.75\textwidth, height=0.25\textwidth]{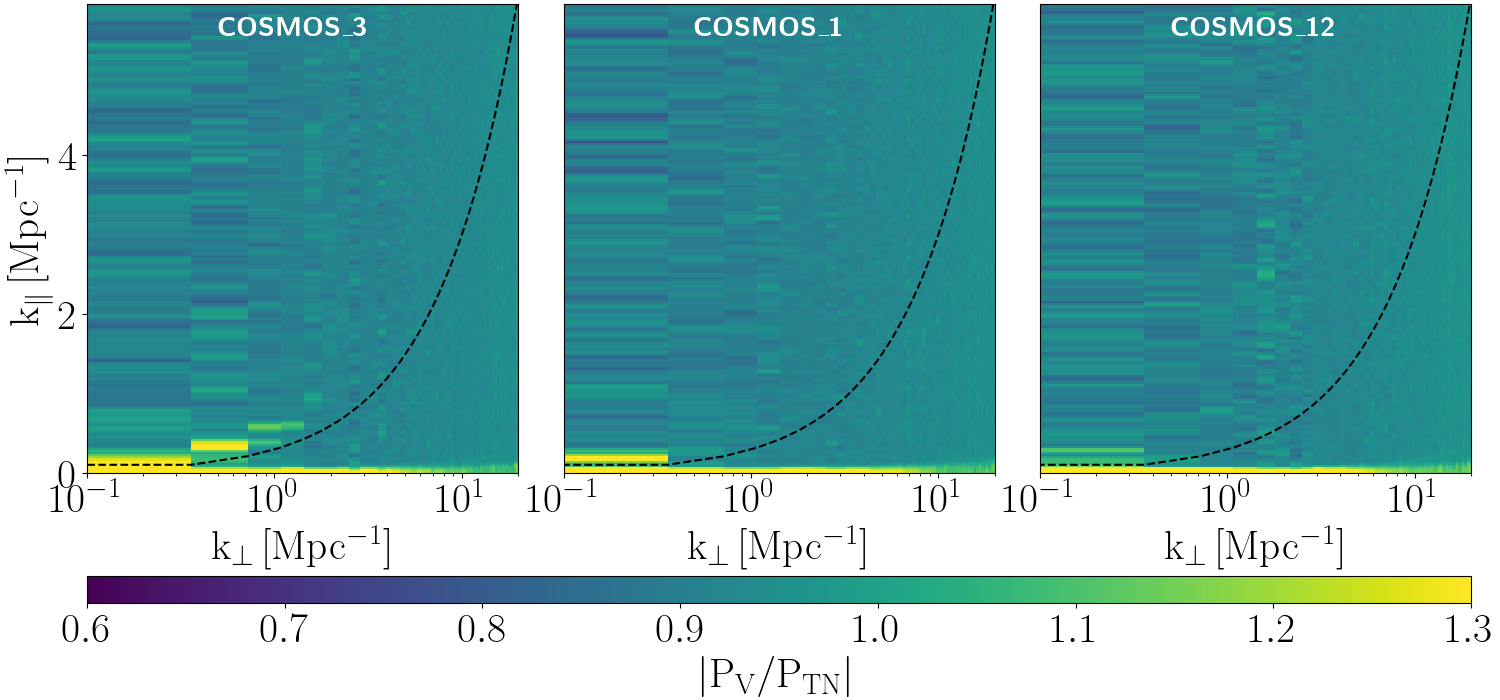}
\includegraphics[width=0.75\textwidth, height=0.25\textwidth]{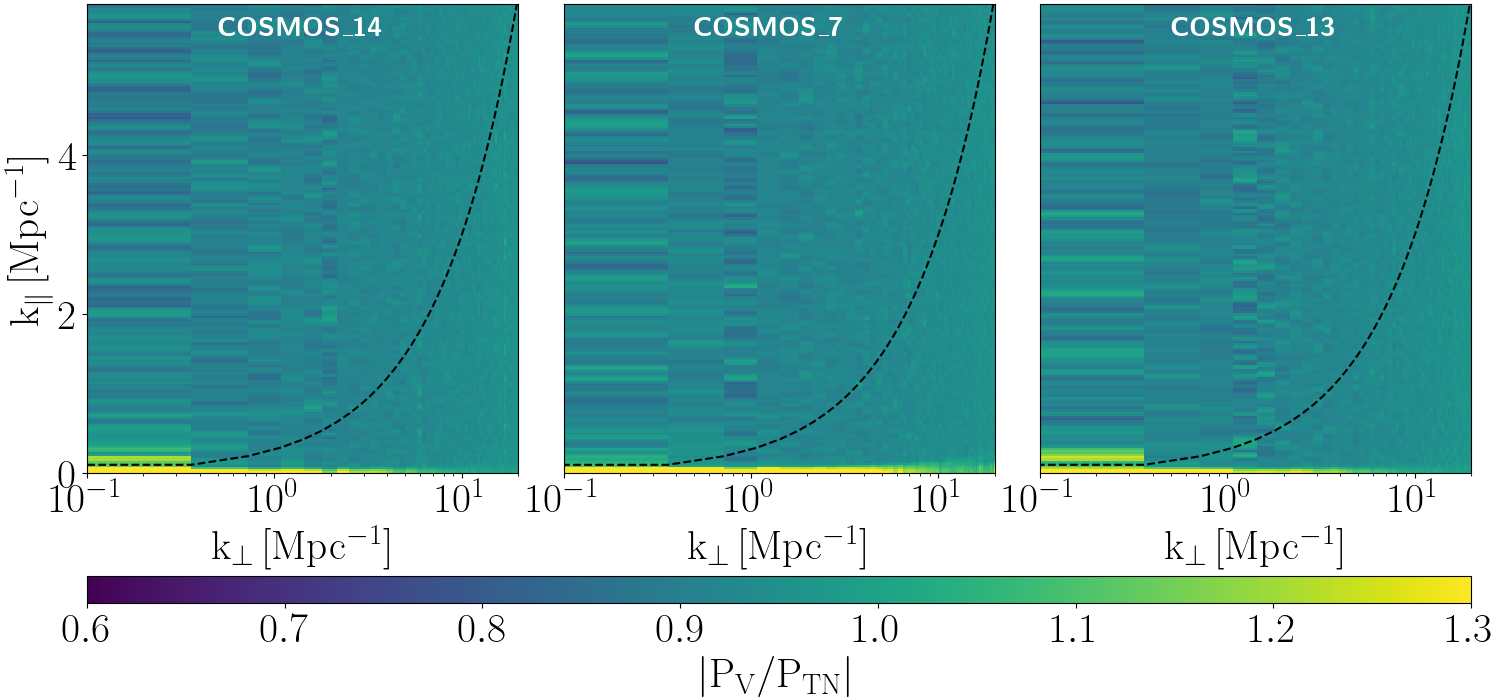}
\caption{Ratio of the cylindrical power spectrum Stokes V to simulated thermal noise power spectrum for individual pointings. An excess power in Stokes V is seen, particularly at the smallest ${k}_{\perp}$ modes, similar to the foreground wedge in Stokes I.}
\label{stokesV_ratio}
\end{figure*}

Since the circular polarization of the sky is assumed minimal, the Stokes V power spectrum is a good approximation to the noise power for each pointing. The ratio of the simulated thermal noise power to the Stokes V cylindrical power spectrum per pointing is shown in Figure \ref{stokesV_ratio}. 

There is a structure spread across the lowest part of the $k_{\perp}$-$k_{\parallel}$ plane, similar to the foreground wedge found in Stokes I. The ratio of the two powers is $\sim$1 outside this region, making the Stokes V and noise powers consistent. The excess is an indication of power leakage into Stokes V for MIGHTEE COSMOS data. It is seen from Figure \ref{stokesV_ratio} that the feature is present across all pointings, showing maximum excess in Stokes V power in the lowest part of $k_{\perp}$ - $k_{\parallel}$ plane (i.e. low delays and very short baselines). The enhanced noise and use of continuum-subtracted data make the feature disappear at higher k$_\perp$. The leakage appears in modes excluded in the power spectrum estimation using foreground avoidance. The modes being used in our final result are reasonably noise-like. Thus, this systematic feature does not affect our final results. Detailed studies into the causes and extent of the leakage for the full MIGHTEE data and its effects on including contaminated modes are deferred to future works.

\bsp	
\label{lastpage}
\end{document}